\RequirePackage{fix-cm}

\documentclass[twocolumn]{svjour3}          
\smartqed  
\usepackage{subcaption}
\usepackage{graphicx}
\usepackage[export]{adjustbox}
\usepackage[ruled,vlined]{algorithm2e}
\usepackage{multirow}
\usepackage[noend]{algpseudocode}
\usepackage{subcaption}
\usepackage{xcolor}
\usepackage{hyperref}
\usepackage{comment}
\usepackage{multicol}
\usepackage{xcolor}
\usepackage{MnSymbol}
\usepackage{float}
\usepackage{tikz}
\usepackage{authblk}
\usepackage{amsfonts}

\begin{document}

	\title{Efficient and Accurate Pneumonia Detection Using a Novel Multi-Scale Transformer Approach}
	\author{}
	

	\author{Alireza Saber\textsuperscript{1} \and
		Amirreza Fateh\textsuperscript{2,}\and
		Pouria Parhami\textsuperscript{1} \and
		Alimohammad Siahkarzadeh\textsuperscript{1} \and
		Mansoor Fateh\textsuperscript{1, *}  \and
		Saideh Ferdowsi\textsuperscript{3, *} 
		}
	
	\institute{
		\textsuperscript{1} Faculty of Computer Engineering, Shahrood University of Technology, Shahrood, Iran \\
		\textsuperscript{2} School of Computer Engineering, Iran University of Science and Technology (IUST), Tehran, Iran \\
		\textsuperscript{3} School of Mathematics, Statistics and Actuarial Science, University of Essex, Colchester, UK \\
		\textsuperscript{*}\emph{Corresponding author} \\
		\email{mansoor\_fateh@shahroodut.ac.ir}\\
		\email{s.ferdowsi@essex.ac.uk} }
	\maketitle
	
	\begin{abstract}
		
		Pneumonia, a prevalent respiratory infection, remains a leading cause of morbidity and mortality worldwide, particularly among vulnerable populations. Chest X-rays serve as a primary tool for pneumonia detection; however, variations in imaging conditions and subtle visual indicators complicate consistent interpretation. Automated tools can enhance traditional methods by improving diagnostic reliability and supporting clinical decision-making. In this study, we propose a novel multi-scale transformer approach for pneumonia detection that integrates lung segmentation and classification into a unified framework. Our method introduces a lightweight transformer-enhanced TransUNet for precise lung segmentation, achieving a Dice score of 95.68\% on the "Chest X-ray Masks and Labels" dataset with fewer parameters than traditional transformers. For classification, we employ pre-trained ResNet models (ResNet-50 and ResNet-101) to extract multi-scale feature maps, which are then processed through Convolutional Residual Attention Module and modified transformer module to enhance pneumonia detection. This integration of multi-scale feature extraction and lightweight attention mechanisms ensures robust performance, making our method suitable for resource-constrained clinical environments. Our approach achieves 93.75\% accuracy on the "Kermany" dataset and 96.04\% accuracy on the "Cohen" dataset, outperforming existing methods while maintaining computational efficiency.  \color{red}https://github.com/amirrezafateh/Multi-Scale-Transformer-Pneumonia
		\color{black}
		
	\end{abstract}

	\section{Introduction}
	\label{intro}

	Pneumonia is a serious respiratory condition that causes inflammation in one or both lungs, leading to symptoms such as fever, cough, and difficulty breathing. This illness is particularly dangerous for young children, accounting for approximately 15\% of mortality in children under the age of five \cite{unicef_pneumonia}. The disease is more prevalent in developing countries, where limited access to healthcare, pollution, overcrowding, and poor living conditions exacerbate its effects \cite{who_pneumonia_in_children}.

	Early and accurate diagnosis is essential for effective treatment; however, pneumonia can be challenging to identify due to its similarity to other lung diseases \cite{health_poverty_action_poverty_and_health}. Chest X-rays are commonly used for diagnosis due to their cost-effectiveness and non-invasive nature \cite{radiologyinfo_chest_xray}. Nevertheless, the interpretation of these images can vary significantly, underscoring the necessity for consistent and automated diagnostic tools.

	Recent advancements in deep learning, particularly in Convolutional Neural Networks (CNNs), have shown significant promise in improving pneumonia diagnosis from chest X-rays \cite{lin2023enhancing}. These models can analyze medical images with remarkable precision, often outperforming human radiologists in both consistency and speed. Recent innovations, such as attention mechanisms, have further enhanced diagnostic accuracy \cite{askari2025enhancing}. Additionally, transformers have demonstrated significant potential in medical imaging tasks due to their ability to model long-range dependencies and identify complex patterns \cite{fateh2024msdnet}. These advancements highlight the potential of AI to complement radiologists' expertise and enhance patient outcomes.

	Lung segmentation is a crucial preprocessing step for improving the accuracy of pneumonia detection in chest X-rays. However, this task faces several challenges. First, the presence of artifacts, overlapping anatomical structures, and low contrast in chest X-rays can make it difficult to accurately delineate lung boundaries \cite{song2025multi}. Traditional segmentation models, such as U-Net, while effective, often struggle to capture fine-grained details and contextual information, resulting in suboptimal performance in complex cases \cite{junia2024deep}. Additionally, the high variability in lung shapes and sizes across patients further complicates the segmentation process. Although transformer-based models have shown promise in addressing these issues, they often come with high computational costs and large parameter counts, making them unsuitable for resource-constrained environments.
	
	The classification of pneumonia from chest X-rays also presents significant challenges. First, subtle visual indicators of pneumonia, such as small opacities or localized consolidations, can be easily missed by traditional CNN-based models like ResNet and DenseNet, which primarily focus on local features. While transformer-based models excel at capturing global contextual information, they require large amounts of labeled data and extensive computational resources, limiting their practicality in real-world clinical settings \cite{shavkatovich2025binary,buriboev2024concatenated}. Moreover, the domain shift between pre-trained models (e.g., those trained on ImageNet) and medical imaging datasets often results in suboptimal performance, necessitating advanced techniques like transfer learning and domain adaptation \cite{kanwal2024current}. These challenges highlight the need for a computationally efficient and accurate model that can effectively leverage both local and global features for pneumonia diagnosis.

	To address these challenges, we propose an innovative approach that leverages deep learning through an integrated lightweight transformer, significantly reducing the number of parameters compared to traditional transformers while maintaining lower model complexity. Our method begins with lung segmentation using a TransUNet model, which integrates transformer-based attention mechanisms into the U-Net architecture. The TransUNet model is trained on the "Chest X-ray Masks and Labels" dataset \cite{candemir2013lung,jaeger2013automatic} to accurately segment lung regions in the images. Once trained, this pre-trained model is used with frozen weights to predict lung masks for our target datasets, "Kermany" \cite{kermany2018labeled} and "Cohen" \cite{cohen2020covid}. This segmentation step isolates the lung regions, thereby enhancing the subsequent classification task.
	
	For classification, we utilize pre-trained ResNet models, specifically ResNet-50 and ResNet-101, as the foundation for feature extraction. By extracting multi-scale feature maps from various stages of the ResNet models, we can leverage multiple feature spaces, which enhances the accuracy of our detection. This is achieved through a customized transformer module that employs a cross-attention mechanism, allowing us to make decisions based on more than one feature space. This transformer has been optimized to minimize the number of parameters while preserving performance. By concentrating on the relevant lung regions and integrating multi-scale information, our approach aims to achieve high diagnostic accuracy for pneumonia detection. This architecture reduces the computational load and ensures robust and reliable performance, making it suitable for deployment in resource-limited settings.

	Our proposed method offers the following key contributions:
	\begin{itemize}
		
		\item Development of a novel transformer structure that significantly reduces complexity compared to traditional transformer-based models while maintaining high performance.
		\item Introduction of a novel TransUNet architecture for the segmentation task, achieving a Dice score of 95.68\% on the "Chest X-ray Masks and Labels" dataset. 
		\item Introducing a Convolutional Residual Attention Module (CRAM) that enriches feature representation by integrating multi-layer residual learning with lightweight attention mechanisms.
		\item Incorporation of multi-scale feature extraction, enabling enhanced performance through the utilization
		of multiple feature spaces.
		\item Achieving high accuracy rates of 93.75\% on the "Kermany" dataset and 96.04\% on the "Cohen" dataset.
	\end{itemize}
	
	\section{Related Work}
	\label{relate}

	In recent years, the focus of research on diagnosing and categorizing lung diseases, including pneumonia, through medical imaging has intensified, driven by advances in machine learning and deep learning technologies \cite{jennifer2023neutrosophic}. Precisely segmenting lung areas in chest X-ray (CXR) images is essential for reliable disease identification and thorough analysis. This section examines deep learning techniques for segmenting and diagnosing lung diseases in chest X-ray (CXR) images. For the segmentation task, we focus on the U-Net architecture and its variations, including attention mechanisms and transformer blocks, which have significantly advanced lung disease segmentation. In the classification task, we categorize approaches into basic deep learning models, transfer learning, fine-tuning, and custom models, emphasizing how these advanced techniques have progressively improved diagnostic outcomes.
	
	\subsection{Segmentation}

	\subsubsection{U-Net for CXR Segmentation}
	The U-Net architecture, with its encoder-decoder structure and skip connections, has occurred as a leading method for CXR segmentation. This setup, which captures high-level semantic information and low-level details, is crucial for accurately outlining lung boundaries. Studies have consistently shown U-Net's effectiveness in segmenting lung regions with high accuracy, a factor that significantly comforts the potential of this technology in improving diagnostic outcomes \cite{fateh2023persian,fateh2022providing}. 
	U-Net, introduced by Ronneberger et al., has become a fundamental tool in medical image segmentation \cite{ronneberger2015u}. 
	Additionally, Liu et al. \cite{liu2022automatic} employed a pre-trained EfficientNet-B4 and developed an enhanced version of U-Net for identifying and segmenting lung regions.

	However, traditional U-Net architectures face several limitations that impede their effectiveness in complex segmentation tasks. These limitations include the inability to leverage multi-scale information, which is essential for capturing fine-grained details, and difficulties in extracting rich contextual information, particularly for small or complex anatomical structures \cite{huang2020unet}. Furthermore, the simple skip connections in U-Net may transfer irrelevant or noisy features, leading to ambiguity in feature representation and reduced segmentation accuracy \cite{jha2020doubleu}. These challenges are especially problematic in chest X-rays, where overlapping structures and low contrast exacerbate noise.

	\subsubsection{U-Net Enhancements with Transformers}
	To address traditional U-Net limitations, advanced architectures that enhance U-Net's ability to capture multi-scale and contextual information are needed. Recent research has significantly advanced lung segmentation by enhancing the U-Net architecture with attention mechanisms. Oktay et al. \cite{oktay2018attention} introduced mechanisms that enable the model to concentrate on the most crucial areas within chest X-rays using Attention Gates (AGs). This innovation enhances segmentation accuracy and sensitivity to disease characteristics. 

	Azad et al. and Chen et al. extended the U-Net framework with transformers, demonstrating significant improvements in capturing intricate details and achieving top-tier results in lung segmentation tasks \cite{chen2021transunet}.

	The incorporation of transformer modules has marked a landmark in lung segmentation research. Transformer architectures, known for capturing long-range dependencies and contextual information from text, have been successfully integrated into U-Net variants, leading to notable improvements in segmentation accuracy. For instance, Chen et al. \cite{chen2023cotrfuse} created a hybrid CNN-Transformer model for medical image segmentation, merging the strengths of CNNs and transformers to enhance accuracy and robustness in lung tissue segmentation. 

	\subsection{Classification}
	
	\subsubsection{Classical Approaches for CXR Classification}
	
	Early methods for classifying chest X-ray (CXR) images primarily depended on traditional machine learning techniques, employing classifiers such as Support Vector Machines (SVM), K-nearest Neighbors (k-NN), and Random Forests. For example, Stokes et al. used logistic regression, decision trees, and SVM to categorize patients' clinical data into bronchitis or pneumonia, with decision trees yielding the highest recall value of 80\% and an AUC of 93\% \cite{stokes2021machine}. 
	Chandra et al. used a multi-layer perceptron (MLP) to segment lung regions from CXR images, reaching an accuracy of 95.39\% \cite{chandra2020pneumonia}. However, these methods, which heavily relied on symptomatic data, had limited accuracy and were evaluated on small datasets \cite{wang2024prediction,fateh2021multilingual}.

	\subsubsection{Deep Learning Models}
	
	The beginning of deep learning, especially Convolutional Neural Networks (CNNs), has significantly transformed medical image analysis by providing superior accuracy and robustness \cite{allioui2022multi}. For instance, Stephen et al. designed a custom CNN model from scratch, achieving a training accuracy of 95.31\% and a validation accuracy of 93.73\% \cite{stephen2019efficient}. Similarly, Sharma et al. created a straightforward CNN architecture that reached a 90.68\% accuracy rate on the "Kermany" dataset using data augmentation \cite{kermany2018labeled}. 
	However, relying solely on data augmentation does not introduce substantially new information, restricting the model's ability to learn a wide range of complex patterns from the training data.
	
	\subsubsection{Transfer Learning}
	Pre-trained CNNs have become the standard for image classification tasks, including CXR analysis. These models leverage large datasets and transfer learning to enhance performance on specific medical imaging tasks. Transfer learning, where pre-trained models are adapted and refined for new, specific tasks, has achieved significant results. For instance, Rajpurkar et al. utilized DenseNet-121 on the ChestX-ray8 dataset, comprising 112,150 frontal CXR images, achieving an F1-score of 76.8\%. This study highlighted the potential of transfer learning in medical image classification \cite{rajpurkar2017chexnet}. 

	\subsubsection{Ensemble Approaches}
	Ensemble learning, which combines the outputs of multiple CNN models, has shown considerable promise. For instance, Ukwuoma et al. \cite{ukwuoma2023hybrid} proposed two ensemble methods: ensemble group A (DenseNet201, VGG16, and GoogleNet) and ensemble group B (DenseNet201, InceptionResNetV2, and Xception). These models, followed by a self-attention layer and a multi-layer perceptron (MLP) for disease identification, achieved 97.22\% accuracy for binary classification, and 97.2\% and 96.4\% for multi-class classification, respectively. 
	Jaiswal et al. \cite{jaiswal2019identifying} used a mask region-based CNN for pneumonia detection through segmentation, employing an ensemble of ResNet-50 and ResNet-101 for image thresholding. 

	Despite their success, pre-trained models such as ResNet have inherent limitations. While ResNet models are powerful, they often struggle to independently capture all the discriminative features required for specific tasks, particularly in complex medical imaging scenarios like pneumonia detection \cite{jaiswal2019identifying}. This limitation is evident in studies where ResNet architectures require complementary support from other models or advanced techniques, such as snapshot ensembling and weighted averaging, to achieve optimal performance \cite{gabruseva2020deep}. Furthermore, ResNet's reliance on local feature extraction through convolutional layers can hinder its ability to model long-range dependencies and global contextual information, which are crucial for accurate classification in medical images \cite{chen2021transunet}. These shortcomings highlight the need for more robust frameworks, such as transformers, which excel at capturing global context and intricate patterns, thereby addressing the limitations of traditional CNN-based models like ResNet.


	\subsubsection{Transformers}
	Recent advancements in medical image classification have harnessed transformer architectures alongside deep learning, yielding impressive outcomes \cite{Gholamiije25}. Wang et al. \cite{wang2021transpath} unveiled TransPath, a hybrid model merging CNN and transformer architectures, highlighting the potential of such integrations. They proved the efficacy of self-supervised pretraining on extensive datasets like TCGA and PAIP, followed by fine-tuning on specific medical image datasets, resulting in solid performance: 89.68\% accuracy on MHIST, 95.85\% on NCT-CRC-HE, and 89.91\% on PatchCamelyon. Transformer-based models have garnered attention for their capacity to capture long-range dependencies in images. Wu et al. \cite{wu2022swin} introduced a Swin Transformer-based model for pulmonary nodule classification, successfully adapting the architecture to the smaller scale of medical image datasets and achieving significant results. 

	In recent years, transformer-based models have continued to evolve, with a particular emphasis on improving efficiency and accuracy in medical imaging tasks. The Swin Transformer V2 \cite{liu2022swin} has emerged as a powerful architecture for various medical imaging tasks, including pneumonia detection. It achieves superior performance by leveraging hierarchical feature extraction and shifted window mechanisms, which allow it to capture both local and global patterns in chest X-rays. In a recent study, the Swin Transformer V2 achieved an accuracy of 98.6\% on a diverse chest X-ray dataset, outperforming traditional CNNs like ResNet and DenseNet \cite{mishra2024empowering}. This highlights its potential for clinical applications where high diagnostic accuracy is essential.

	\begin{figure*}[h]
		\centering
		\includegraphics[width=\linewidth]{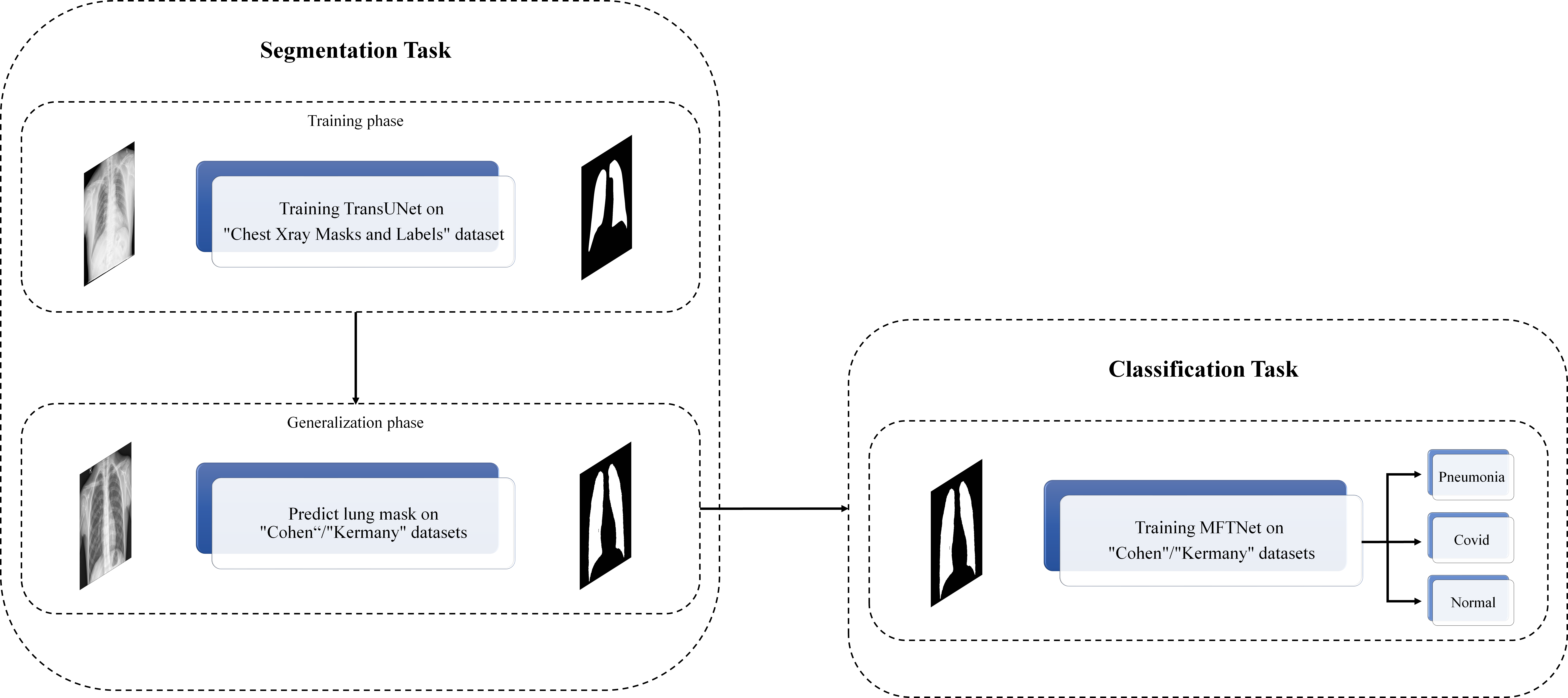}
		
		\caption{The block diagram of the proposed method}

		\label{fig:block_dig}
	\end{figure*}
	
	Hybrid architectures that combine CNNs and transformers have demonstrated remarkable success in pneumonia detection. For instance, a hybrid model integrating ResNet34 with a Multi-Axis Vision Transformer achieved a state-of-the-art accuracy of 94.87\% on the Kaggle pediatric pneumonia dataset. This model leverages the local feature extraction capabilities of CNNs and the global context modeling of transformers, resulting in fewer misclassifications and improved robustness \cite{angara2024novel}. 

	While transformer-based models have shown significant promise in medical image classification, they face several limitations when applied to pneumonia detection. First, many existing transformer architectures, such as Vision Transformers (ViTs) and Swin Transformers, require large computational resources and extensive training data, making them unsuitable for resource-constrained clinical environments \cite{liu2022swin}. Additionally, these models often struggle to effectively combine local feature extraction (a strength of CNNs) with global context modeling (a strength of transformers), leading to suboptimal performance in tasks like pneumonia detection, where both fine-grained details and global patterns are critical \cite{angara2024novel}. Furthermore, the high parameter counts and complexity of traditional transformers can result in longer training times and higher hardware requirements, limiting their practicality in real-world applications \cite{mustapha2025enhanced}.

	\section{Proposed method}
	\label{proposed}

	\begin{figure*}[h]
		\centering
		\includegraphics[width=\linewidth]{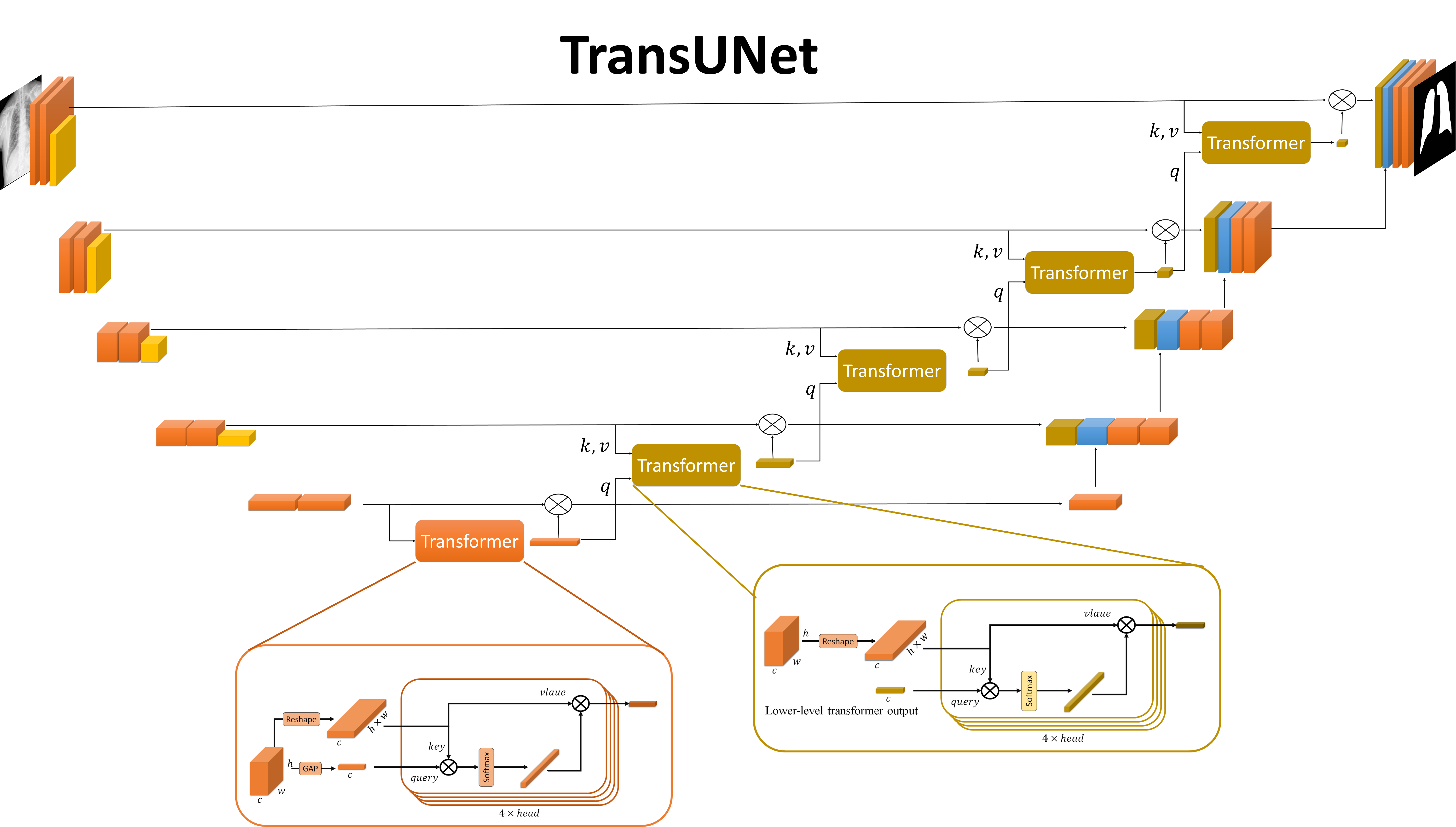}
		
		\caption{The TransUnet Architecture}

		\label{fig:transunet}
	\end{figure*}
	
	\subsection{Overview}
	
	In this study, we propose a novel approach for segmentation and classification of Pneumonia Chest X-ray images by leveraging the power of deep learning and transformer-based attention mechanisms. Our method utilizes pre-trained ResNet models, specifically ResNet-50 and ResNet-101, as the backbone for feature extraction. These models are well-known for their ability to capture intricate patterns and features in images due to their deep architecture and residual connections.

	Our approach begins with a segmentation step where we employ a TransUNet model, which integrates transformer-based attention mechanisms into the popular U-Net architecture. This model is trained on "Chest Xray Masks and Labels" dataset \cite{candemir2013lung,jaeger2013automatic} to accurately segment lung regions in the images. By predicting masks for "Cohen" dataset \cite{cohen2020covid} using this pre-trained TransUNet, we can isolate the regions of interest, enhancing the subsequent classification task.	The segmentation step provides us with precise lung masks, ensuring that our classification model focuses on the relevant areas of the X-ray images. This preprocessing step is crucial for improving the overall accuracy of the system by reducing background noise and irrelevant features.
	
	Our classification approach extracts multi-scale feature maps from three key stages of the ResNet models: the outputs of Block 2, Block 3, and Block 4. These stages provide a rich set of features at different scales, which are crucial for accurately identifying Pneumonia in chest X-rays. The extracted features are first refined through CRAM that enhance discriminative features through dual attention mechanisms and residual learning. These enhanced feature maps are then processed through a specialized transformer module that employs an attention mechanism, further refining the representation by allowing the network to focus on the most relevant parts of the image.
	
	After the attention processing, the feature maps are concatenated to form a comprehensive representation of the input image. This combined feature map is subsequently fed into fully connected layers to perform the final classification. The overall architecture is designed to effectively integrate multi-scale information and attention mechanisms, thereby improving the classification accuracy. The block diagram of the proposed method is illustrated in Figure \ref{fig:block_dig}.

	\subsection{Segmentation task}
	
	In our proposed method, the segmentation task is pivotal for isolating lung regions in chest X-ray images, thereby enhancing the accuracy of pneumonia classification. For this purpose, we have designed a TransUNet model, which uniquely combines the strengths of the U-Net architecture with advanced techniques.
	
	\subsubsection{TransUNet Architecture}

	The TransUNet architecture can be divided into three main components: the encoder, the bottleneck, and the decoder. The encoder consists of a series of convolutional layers designed to capture hierarchical features from the input image. Each stage of the encoder includes a double convolution block, which performs two consecutive convolutions followed by batch normalization and ReLU activation. This setup helps in learning complex features at multiple levels. The encoder progressively reduces the spatial dimensions while increasing the depth of the feature maps through max-pooling operations. 
	
	At the bottleneck stage, the most abstract features of the input image are captured. This layer consists of a double convolution block. The bottleneck also incorporates an embedding layer and a positional encoding mechanism, which prepare the feature maps for the subsequent transformer module. The detailed structure and function of the transformer module will be discussed later in the classification subsection.
	
	Also, the transformer modules integrate into each skip connection between the encoder and decoder. These transformers enhance the model's ability to capture global contextual information at each resolution level. In this design, the query for each transformer's attention mechanism is derived from the output of the transformer at the preceding, lower level. The transformer's output is then element-wise multiplied with the original skip connection feature maps before being passed to the decoder. This ensures that the information passed to the decoder not only retains local details but also incorporates refined global representations.
	
	The decoder reconstructs the segmented output by progressively upsampling the feature maps and concatenating them with the outputs of the transformer-augmented skip connections. Each upsampling step is followed by a double convolution block to refine the features and reduce the number of channels. This structure allows the decoder to restore the spatial resolution of the feature maps while retaining the detailed information captured by the encoder. Finally, a convolutional layer with a single output channel generates the segmentation mask for the lung regions.
	
	\subsubsection{Training the TransUNet Model}
	
	\begin{figure*}[h]
		\centering
		\includegraphics[width=\linewidth]{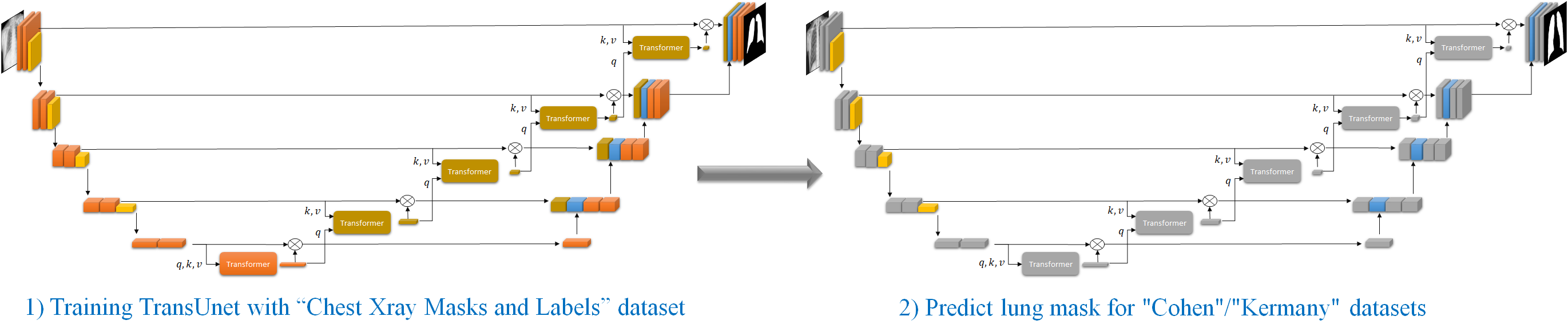}
		
		\caption{Applying trained TransUNet to predict lung masks on "Cohen"/"Kermay" datasets}

		\label{fig:predicted_mask}
	\end{figure*}

	Initially, we resize all images to 512x512 pixels. The TransUNet model is trained on the "Chest Xray Masks and Labels" dataset, which provides paired X-ray images and corresponding lung masks. By training on this dataset, the TransUNet model learns to accurately segment lung regions in chest X-rays, ensuring that the subsequent classification step focuses on the relevant areas, thereby improving the overall accuracy of the system.
	
	By effectively combining the robust feature extraction capabilities of the U-Net architecture with advanced processing techniques, the TransUNet model provides a powerful solution for the segmentation task in our proposed method.

	\subsubsection{Applying the Trained TransUNet to Cohen/Kermany Datasets}
	As shown in Figure \ref{fig:predicted_mask}, after successfully training the TransUNet model on the "Chest Xray Masks and Labels" dataset, we utilize this pre-trained segmentation model to predict lung masks for the "Cohen"/"Kermany" datasets. This step is crucial for enhancing the accuracy of the subsequent classification task by focusing on the lung regions within the X-ray images.
	
	The "Cohen"/"Kermany" datasets, which contain chest X-ray images, requires preprocessing to ensure that our classification model focuses on the most relevant regions. To achieve this, we apply the trained TransUNet model to segment the lung areas from these images. The "Cohen" dataset is first preprocessed to match the input requirements of the TransUNet model. This involves standardizing the image dimensions and normalizing the pixel values to ensure consistency with the training data used for the TransUNet model.
	
	Using the pre-trained TransUNet model, we generate lung masks for each X-ray image in the "Cohen"/"Kermany" datasets. The segmentation model outputs binary masks that highlight the lung regions while suppressing the background. By utilizing the pre-trained TransUNet model to segment the lung regions in the "Cohen" dataset, we effectively preprocess the data to improve the performance of our classification model. This segmentation step filters out noise and irrelevant features, allowing the classifier to concentrate on the lung areas, thereby enhancing the overall accuracy and robustness of our proposed method.

	\subsection{classification task}
\begin{figure*}[h]
		\centering
		\includegraphics[width=\linewidth]{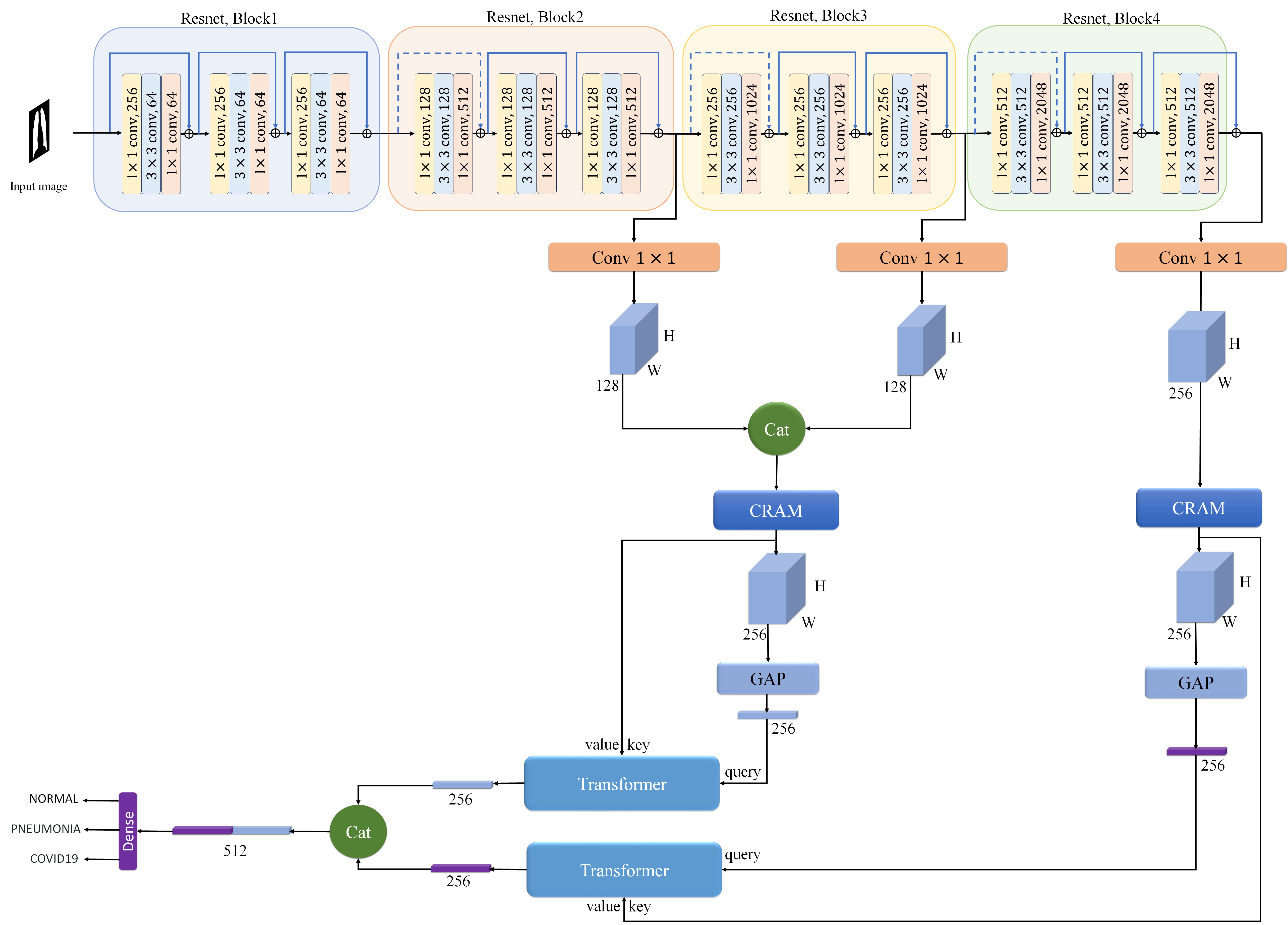}
		
		\caption{The overview of the proposed method on classification task}

		\label{fig:classification_task}
	\end{figure*}

	Following the segmentation of lung regions using the TransUNet model, the next step in our proposed method is the classification task. This task involves accurately identifying the presence of pneumonia or covid19 in the preprocessed chest X-ray images. By focusing on the lung regions isolated during the segmentation phase, we enhance the classification model's ability to detect relevant features indicative of pneumonia or covid19, thereby improving diagnostic accuracy. The overview of the proposed method on classification task is shown in Figure \ref{fig:classification_task}.
	
	\subsubsection{Backbone}
	
	The backbone of our proposed method utilizes a pre-trained ResNet model, specifically ResNet-50 or ResNet-101, to extract multi-scale feature maps from the input chest X-ray images. Initially, we resize all images to 512x512 pixels. We focus on the outputs from Block 2, Block 3, and Block 4 of the ResNet, denoted as $B^2$, $B^3$, and $B^4$ respectively. Each of these blocks provides feature maps of size $(c,h,w)$, where $h=w=64$ The channels $c$ for these blocks are 512, 1024, and 2048 respectively.

	The selection of Block 2, Block 3, and Block 4 is motivated by their unique contributions to the classification task. Block 2 captures low-level features such as edges and textures, which are essential for identifying subtle patterns in the lung regions. Block 3 extracts mid-level features, including more complex structures like lung lobes and localized consolidations, which are critical for detecting early signs of pneumonia. Block 4 provides high-level semantic features, such as global contextual information and disease-specific patterns, which are necessary for accurate classification. By combining these multi-scale features, our model can effectively capture both fine-grained details and global context, leading to improved diagnostic performance.

	To handle the complexity and standardize the feature maps for subsequent processing, we apply 1x1 convolution operations to reduce the number of channels. Specifically, for the output of Block 4 ($B^4$), as shown in Equation \ref{b4}, we reduce the channels to 64 using a 1x1 convolution.
	\begin{equation}
		B'^4=C_{1\times1}(B^4)
		\label{b4}
	\end{equation}
	where $C_{1\times1}$ denotes the 1x1 convolution operation.
	
	For the outputs of Blocks 2 and 3 ($B^2$ and $B^3$), as shown in Equation \ref{b2} and Equation \ref{b3}, we use separate 1x1 convolutions to reduce the number of channels for each to 32.
	\begin{equation}
		B'^2=C_{1\times1}(B^2)
		\label{b2}
	\end{equation}
	\begin{equation}
		B'^3=C_{1\times1}(B^3)
		\label{b3}
	\end{equation}
	where $C_{1\times1}$ in each equation indicates a reduction in the number of channels to 32.
	
	After reducing the channels, we concatenate the feature maps from Block 2 and Block 3 to form a merged feature map (Equation \ref{b_merge}).
	\begin{equation}
		B^{merged}=Cat(B'^2,B'^3)
		\label{b_merge}
	\end{equation}
	where $Cat$ denotes the concatenation operation.
	
	Thus, we have two main feature maps with size of (64,64,64) for further processing:
	\begin{itemize}
		\item $B'^4$
		\item $B^{merged}$
	\end{itemize}
	
	\begin{figure*}[h]
		\centering
		\includegraphics[width=0.8\linewidth]{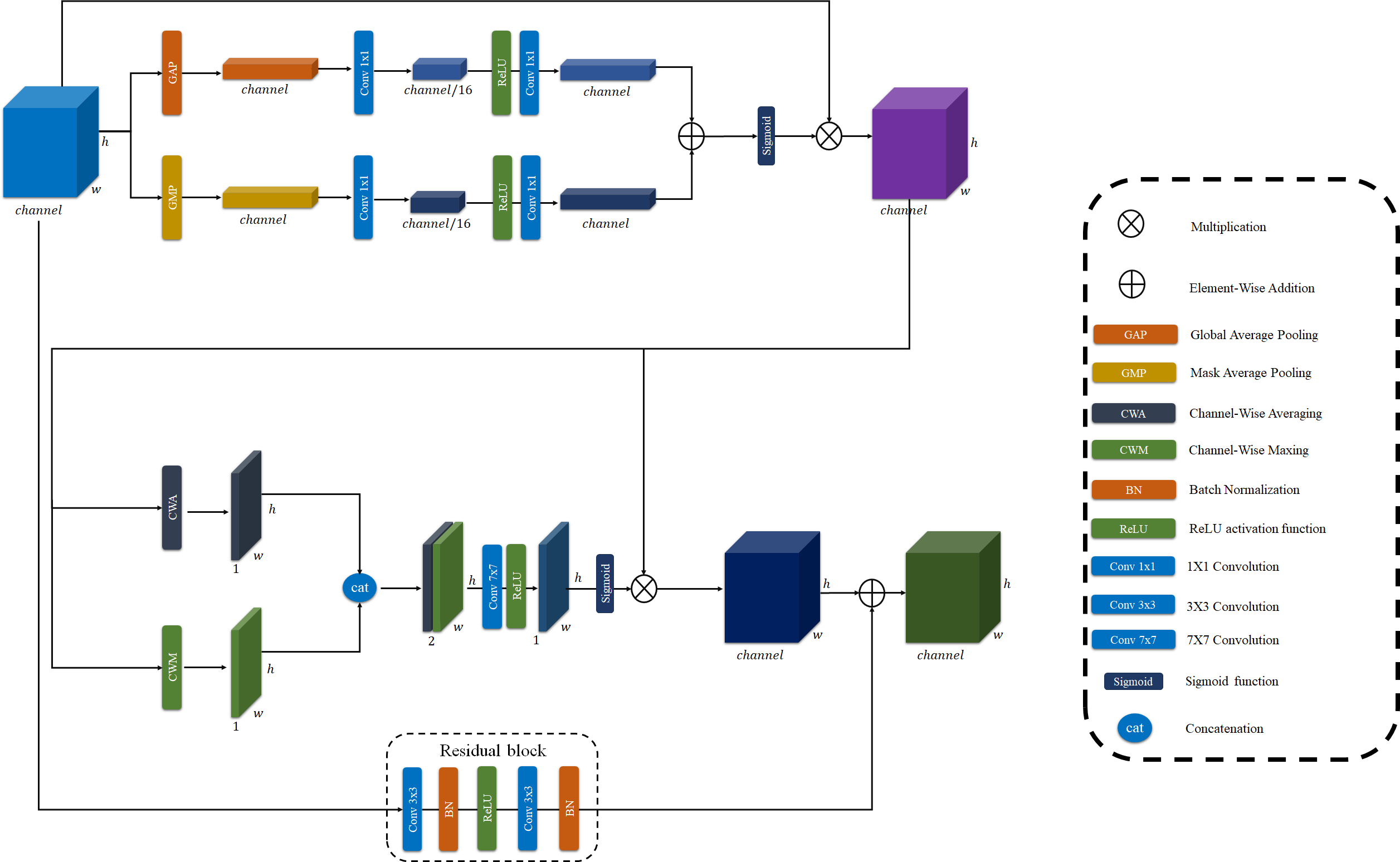}
		
		\caption{The overview of CRAM}

		\label{fig:cram}
	\end{figure*}

	\subsubsection{Convolutional Residual Attention Module (CRAM)}
	
	To further refine the multi-scale feature representations extracted from the ResNet backbone, we propose CRAM. This module operates on two parallel pathways to enhance feature discriminability while preserving spatial relationships. The CRAM is applied to both feature streams: the high-level semantic features from Block 4 ($B'^4$) and the merged multi-scale features from Blocks 2 and 3 ($B^{merged}$).
	
	As shown in Figure \ref{fig:cram}, the CRAM architecture consists of two complementary components working together. For an input feature map $X \in \mathbb{R}^{C \times H \times W}$, the module processes it through both pathways simultaneously.
	
	\textbf{1. Dual-Attention Mechanism:} This module incorporates a sophisticated attention mechanism that operates through two distinct dimensions.
	
	\textbf{Channel Attention:} This component generates a channel attention vector denoted as $F_{CA}$  that emphasizes informative feature channels. The computation is defined as Equation \ref{eq:ac}. We also define a Sequential Convolution (SC) which have been used on both GAP and GMP that showed in Equation \ref{eq:sc}.
	
	\begin{equation}
		\label{eq:ac}
		F_{CA}(x) = \sigma (GAP(SC) + GMP(SC(x)))\times x
	\end{equation}
	\begin{equation}
		\label{eq:sc}
		SC = conv_{1\times1}(Relu(conv_{1\times1}(x)))
	\end{equation}

	where $\text{GAP}$ and $\text{GMP}$ represent global average pooling and global max pooling operations respectively, SC denotes two-layer 1×1 convolutional network with ReLU activation that reduces channels to $\frac{1}{16}$ and then expands back to original channels. $\sigma$ is the sigmoid function. Finally the channel-refined features are obtained through element-wise multiplication.
	
	\textbf{Spatial Attention:} This component focuses on identifying spatially significant regions within the feature maps. As shown in Equation \ref{eq:sa} and Equation \ref{eq:cat_sa}, it computes both average-pooled and max-pooled features across the channel dimension, concatenates them, and processes them through a convolutional layer with a sigmoid activation to produce spatial attention maps that highlight important spatial locations.

	\begin{equation}
		\label{eq:sa}
		F_{SA}(x) = \sigma (Relu(conv_{7\times7}(cat_{SA})) 
	\end{equation}
	
	\begin{equation}
		\label{eq:cat_sa}
		cat_{SA} = cat(\text{CWA}(F_{CA}(x)),\text{CWM}(F_{CA}(x)))
	\end{equation}
	
	where $F_{CA}$ denotes the output of channel attention part, CWA and CWM represent Channel-Wise Averaging and Channel-Wise Maxing
	respectively, $cat$ denotes concatenation.  
	
	The final output of the dual-attention mechanism is the element-wise product of the spatially and channel-wise refined features, as formalized in Equation \ref{eq:at}.

	\begin{equation}
		\label{eq:at}
		\text{DAM}(X) = F_{SA}(x) \times F_{CA}(x)
	\end{equation}

	\begin{figure*}[h]
		\centering
		\includegraphics[width=0.8\linewidth]{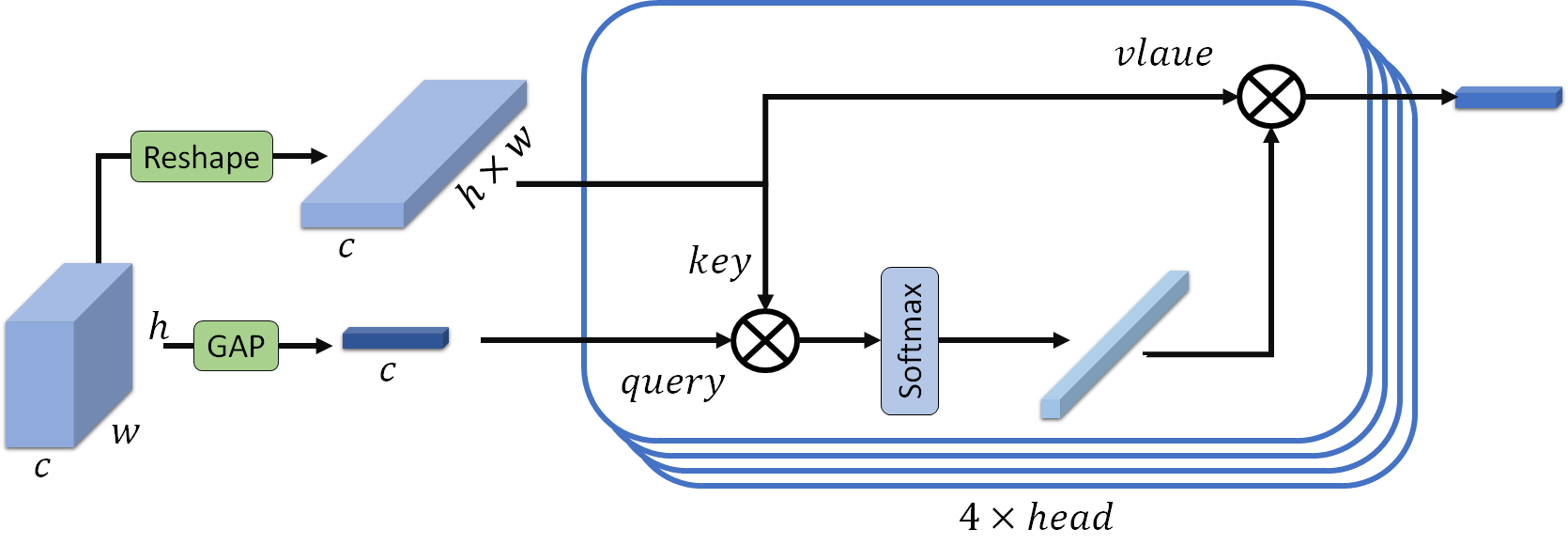}
		
		\caption{The overview of transformer}

		\label{fig:trans}
	\end{figure*}
	
	\textbf{2. Parallel Convolutional Pathway:} A complementary convolutional branch, formalized in Equation \ref{eq:re}, transforms the input through two successive 3×3 convolutional layers with batch normalization and ReLU non-linearities. This pathway facilitates the learning of complex feature mappings while ensuring stable gradient flow via skip connections.
	
	\begin{equation}
		\label{eq:re}
		\text{Re}(X) = \text{BN}(\text{Conv}_{3\times3}(\text{ReLU}(\text{BN}(\text{Conv}_{3\times3}(X)))))
	\end{equation}
	where BN denotes Batch Normalization.
	
	The outputs from both pathways are integrated through element-wise addition to produce the final enhanced representation, as formalized in Equation \ref{eq:cram}. This composite feature enhancement is subsequently applied to both multi-scale feature streams in our architecture, yielding the refined feature maps specified in Equation \ref{eq:b_b}.
	
	\begin{equation}
		\label{eq:cram}
		\text{CRAM}(X) = \text{DAM}(X) + \text{Re}(X)
	\end{equation}
	
	\begin{equation}
		\label{eq:b_b}
		\begin{aligned}
			E^{\text{merged}} &= \text{CRAM}(B^{\text{merged}})  \\
			E^4 &= \text{CRAM}(B'^4) 
		\end{aligned}
	\end{equation}
	
	where, $E^{\text{merged}}$ denotes the enhanced multi-scale features from Blocks 2 and 3, while $E^4$ represents the refined high-level semantic features from Block 4.
	

	\subsubsection{Transformer}
	In our proposed method, a transformer is employed to enhance the feature representation obtained from the ResNet backbone, and a similar transformer architecture is used within the TransUNet model. For the classification task, we leverage this transformer to refine the multi-scale feature maps extracted from the ResNet backbone. We begin with two feature maps of size (64,64,64) derived from the ResNet backbone: $E^4$ and $E^{merged}$. Each of these feature maps is fed into a separate transformer, although the structure of the transformers is identical. For simplicity, we will describe the process for $E^4$.
	\paragraph{\textbf{Global Average Pooling (GAP)}}
	We apply global average pooling to $B'^4$ to create a feature vector $V^4$ (Equation \ref{v4}).
	\begin{equation}
		V^4=GAP(E^4)
		\label{v4}
	\end{equation}
	This vector serves as the query for the transformer.
	\paragraph{\textbf{Reshaping for Key and Value}}
	The feature map $E^4$ is reshaped to form the key and value inputs for the transformer. Specifically, as shown in equation \ref{bflat}, $E^4$ is reshaped from $(c,h,w)$ to $(c,h \times w)$.
	\begin{equation}
		E_{flat}^{4} = Reshape(E^4)
		\label{bflat}
	\end{equation}
	
	\paragraph{\textbf{Attention Mechanism}}
	Query($Q$): The query is obtained from the feature vector $V^4$ created by global average pooling.
	
	Key($K$) and Value($V$): Both the key and value are derived from the reshaped feature map $E_{flat}^{4}$
	\paragraph{\textbf{Scaled Dot-Product Attention}}
	
	The attention scores are computed as the dot product of the query and key, followed by a softmax operation to obtain the attention weights (Equation \ref{att}).
	\begin{equation}
		\text{Attention}(Q, K, V) = \text{Softmax}\left(\frac{Q \cdot K^T}{\sqrt{d_k}}\right) \cdot V
		\label{att}
	\end{equation}
	where \( d_k \) is the dimensionality of the key, and \( Q \cdot K^T \) represents the dot product of the query and the transposed key.
	The result is a weighted sum of the value vectors, producing an output feature vector $F^4$ of size equal to the channel dimension.
	\subsubsection{Output Feature Vector}
	The output of the transformer for $E^4$ is a feature vector $F^4$ of size equal to the number of channels (64 in this case).
	
	The same process is applied to $E^{merged}$, resulting in another feature vector $F^{merged}$ of size 64. The entire process is visually represented in Figure \ref{fig:trans}, providing a detailed overview of the transformer's operation on the feature maps.

	\subsubsection{Find correct class}
	After processing the feature maps $F^4$ and $F^{merged}$ through transformers, we concatenate these outputs to form a unified feature representation (Equation \ref{cat}).
	\begin{equation}
		F^{concat} = Cat(F^4,f^{merged})
		\label{cat}
	\end{equation}
	The concatenated feature vector $F^{concat}$ is then flattened into a one-dimensional vector. The flattened feature vector is processed through a dense (fully connected) layer followed by a sigmoid activation function for binary classification.
	\subsubsection{Loss Function}
	We employ binary cross-entropy loss to train the classifier. This loss function measures the discrepancy between predicted probabilities and true labels for binary classification tasks (Equation \ref{loss}).
	\begin{equation}
		\text{Loss} = -\frac{1}{N} \sum_{i=1}^{N} \left[ y_i \cdot \log(\hat{y}_i) + (1 - y_i) \cdot \log(1 - \hat{y}_i) \right]
		\label{loss}
	\end{equation}
	where $N$ is the number of samples, $y_i$ is the true label (0 or 1), and $\hat{y_i}$ is the predicted probability.
	
	\section{Experimental Result}
	\label{exp}
	\subsection{Dataset}
	
	In this research, we utilized several datasets to effectively train and evaluate our models for both segmentation and classification tasks. For training and validating the TransUNet model, we used the "Chest Xray Masks and Labels" dataset \cite{candemir2013lung,jaeger2013automatic}. This dataset contains 714 chest X-ray images, accompanied by their masks. Due to data limitation, our segmentation model was trained on 690 images and their corresponding masks, with 24 images reserved for validation purposes.
	
	For classification task, we used two datasets. COVID-19 Image Data Collection provided by "Cohen" \cite{cohen2020covid}. This dataset comprises a total of 6,432 images, including three classes: Pneumonia, COVID-19, and Normal. The dataset is notably challenging due to its class imbalance and the complexity introduced by the three distinct classes. The distribution of images in the training set is as follows: 3,418 images of Pneumonia, 1,266 images of Normal, and 460 images of COVID-19. Approximately 20\% of the images are allocated for testing.
	
	Pediatric Pneumonia Chest X-ray Dataset provided by Kermany et al. \cite{kermany2018labeled}. This dataset includes 5,856 images, with 5,232 images used for training and the remaining images reserved for testing. The dataset presents a significant challenge due to its class imbalance, with 3,883 images labeled as Pneumonia and 1,349 images as Normal. Additionally, the images in this dataset are from children, who often experience discomfort during X-ray procedures. This discomfort can impact the quality and consistency of the images, and the physiological differences between children and adults add an extra layer of complexity to the classification task. Both datasets contribute valuable and complementary challenges to our classification task, ensuring that our model is robust and capable of handling various real-world scenarios.

	\subsection{Experimental Setting}
	We implemented our proposed method using PyTorch version 1.8.1. In classification task, we utilized pre-trained ResNet-50 and ResNet-101 models, which were kept frozen during training to preserve their learned representations. Notably, the learnable parameters of our method amount to only 2.29 million. The model was trained for 30 epochs, and this process was repeated five times to ensure the robustness of the results. The average of these results was then reported.
	
	The Adam optimizer was employed for training, with a learning rate set to $10^{-5}$. All input images were resized to 512 × 512 pixels, and the batch size was set to 64. All experiments were conducted on an NVIDIA RTX 4090 GPU.

	\subsection{Evaluation Metrics}
	To evaluate the effectiveness of our proposed approach, we use several performance metrics, including:
	
	\textbf{Accuracy}: This metric reflects the ratio of correctly identified instances to the total number of instances. It is determined using Equation \ref{acc}.
	\begin{equation}
		\text{Accuracy} = \frac{\text{TP} + \text{TN}}{\text{TP + TN + FP + FN}}
		\label{acc}
	\end{equation}
	Accuracy offers a broad overview of the classifier's performance but can be deceptive when dealing with imbalanced datasets.
	
	\textbf{Precision}: This metric quantifies the ratio of correctly predicted positive instances to the total predicted positives. It is represented by Equation \ref{per}.
	\begin{equation}
		\text{Precision} = \frac{\text{TP}}{\text{TP} + \text{TN}}
		\label{per}
	\end{equation}
	Precision is especially valuable when the consequence of false positives is significant.

	\textbf{Recall}: Also referred to as Sensitivity or True Positive Rate, Recall measures the ratio of correctly predicted positive instances to the total actual positives. It is expressed by Equation \ref{rec}.
	
	\begin{equation}
		\text{Recall} = \frac{\text{TP}}{\text{TP} + \text{FN}}
		\label{rec}
	\end{equation}
	Recall is vital in situations where the cost of missing a positive instance (false negatives) is high, ensuring that most positive instances are detected.

	\textbf{F1 Score}: The F1 Score represents the harmonic mean of Precision and Recall, offering a balance between these two metrics. It is particularly advantageous when handling imbalanced datasets. The F1 Score is determined using Equation \ref{f1score}.
	\begin{equation}
		\text{F1 Score} = 2\times \frac{\text{Precision}\times\text{Recall}}{\text{Precision}+\text{Recall}}
		\label{f1score}
	\end{equation}
	This score provides a single metric that accounts for both false positives and false negatives, reflecting the classifier’s overall performance.

	\textbf{MCC}: The Matthews Correlation Coefficient (MCC) is a robust metric for evaluating segmentation performance, especially in imbalanced datasets. It considers true positives (TP), true negatives (TN), false positives (FP), and false negatives (FN) to provide a balanced measure of classification quality. As stated in \cite{anbalagan2023analysis}, MCC is particularly useful for tasks where class imbalance is a concern, as it accounts for all four categories of the confusion matrix. MCC is calculated using Equation \ref{mcc}.
	
	\begin{equation}
		\resizebox{0.9\linewidth}{!}{$
			\text{MCC} = \frac{\text{TP} \times \text{TN} - \text{FP} \times \text{FN}}{\sqrt{(\text{TP} + \text{FP}) \times (\text{TP} + \text{FN}) \times (\text{TN} + \text{FP}) \times (\text{TN} + \text{FN})}}
			$}
		\label{mcc}
	\end{equation}

	\textbf{Dice Coefficient}: The Dice Coefficient, also known as the F1-score for segmentation tasks, measures the overlap between the predicted segmentation and the ground truth. It is particularly useful for evaluating the accuracy of region-based segmentation, such as lung segmentation in chest X-rays. The Dice Coefficient is calculated using Equation \ref{dice}:
	\begin{equation}
		\text{Dice} = \frac{2 \times \text{TP}}{2 \times \text{TP} + \text{FP} + \text{FN}}
		\label{dice}
	\end{equation}

	\subsection{Comparison with State-of-the-Art}
	\begin{table*}[h]
		\centering
		
		\caption{Performance on "Chest X-ray Masks and Labels" dataset for segmentation task. Numbers in bold represent the best performance, while underlined values denote the second-best performance.
		}
		
		\renewcommand{\arraystretch}{2}
		\resizebox{2\columnwidth}{!}{
			\begin{tabular}{c|cccccc}
				\hline
				Model & Dice  & Accuracy  & Precision  & Recall  & F1-score  & MCC  \\ \hline
				Unet \cite{ronneberger2015u} & 93.46\% & 96.88\% & 97.41\% & 89.83\% & 93.46\% & 91.55\% \\ 
				RU-Net \cite{alom2018nuclei} & 92.07\% & 96.34\% & \textbf{99.58\%} & 85.61\% & 92.07\% & 90.13\% \\ 
				ResNet34-Unet \cite{lau2020automated} & 93.83\% & 97.06\% & 98.13\% & 89.89\% & 93.83\% & 92.06\% \\ 
				BCDU-Net \cite{azad2019bi} & 94.14\% & 97.20\% & 98.25\% & 90.37\% & 94.14\% & 92.44\% \\ 
				ResBCDUnet \cite{jalali2021resbcdu} & 94.34\% & 97.31\% & \underline{98.89\%} & 90.20\% & 94.34\% & 92.75\% \\ 
				NasNet \cite{zhang2022automatic} & 94.95\% & 97.52\% & 96.55\% & \underline{93.42\%} & 94.95\% & 93.33\% \\ 
				DABT-U-Net \cite{jalali2024dabt} & 95.11\% & 97.64\% & 98.25\% & 92.16\% & 95.11\% & 93.64\% \\ 
				ABANet \cite{rezvani2024abanet} & 95.25\% & 97.71\% & 98.53\% & 92.18\% & 95.25\% & 93.84\% \\ 
				FusionLungNet \cite{rezvani2025fusionlungnet} & \underline{95.29\%} & \underline{97.73\%} & 98.66\% & 92.14\% & \underline{95.29\%} & \underline{93.89\%} \\ 
				Our & \textbf{95.7\%}$\pm$0.5 & \textbf{97.9\%}$\pm$0.5 & 97.5\%$\pm$0.6 & \textbf{93.9\%}$\pm$0.6 & \textbf{95.7\%}$\pm$0.5 & \textbf{94.3\%}$\pm$0.6 \\ \hline
			\end{tabular}
		}
		\label{tb:seg} 
		
	\end{table*}

	\begin{table*}[h]
		\centering
		\caption{Performance on "Cohen" dataset. Numbers in bold represent the best performance, while underlined values denote the second-best performance.
		}
		
		\renewcommand{\arraystretch}{2}
		\resizebox{2\columnwidth}{!}{
			\begin{tabular}{c|cccc}
				\hline
				Models              & Acuracy & Precision & Recall  & F1-score \\ \hline
				Densenet121*         &  87.8\% & 53.9\%    &71.0\%   &61.27\%  \\
				Densenet169*         &  87.1\% & 32.3\%    &65.6\%   &43.28\%  \\
				Densenet201*         &  88.4\% & 51.9\%    &79.0\%   &62.64\%  \\
				Mobilenet\_v2*       &  86.9\% & 33.4\%    &75.0\%   &46.21\%  \\
				ResNet-50*           & 87.1\%  & 38.4\%    &71.0\%   &49.84\%  \\
				ResNet-101*          & 87.9\%  & 33.5\%    &73.0\%   &45.92
				\%  \\
				Goodwinet al. (Ensemble learning) \cite{goodwin2020intra} &89.4\% &53.3\%& 80.0\%  &63.97\%  \\
				Gadza et. al \cite{gazda2021self}& 84.9\% & 77.4\%  & 90.6\%& 83.48\% \\
				
				Zhao et al. (Channel-Attention Capsule) \cite{zhao2023dcacorrcapsnet} & 90.43\% & 90.81\% & 90.43\% & 90.40\% \\
				CNN-based \cite{van2024large}& 92.52\% & -  & -& - \\
				CNN-based \cite{van2024large}& 91.05\% & -  & -& - \\
				Proposed method (ResNet-50 as backbone)  & \textbf{96.04\%}$\pm$0.5   & \textbf{96.70\%}$\pm$0.5  & \textbf{94.90\%}$\pm$0.6    & \textbf{95.77\%}$\pm$0.5\\
				Proposed method (ResNet-101 as backbone) & \underline{95.19\%}$\pm$0.6   &  \underline{95.59\%}$\pm$0.5  & \underline{94.19\%}$\pm$0.6    &   \underline{94.86\%}$\pm$0.6\\
				\hline
			\end{tabular}			
		}
		\label{tb:1} \\
		\footnotesize{\textit{Models marked with "*" have results directly reported from \cite{goodwin2020intra}.}}
	\end{table*}
	
	\begin{table*}[h]
		\centering
		\caption{Performance on "Kermany" dataset. Numbers in bold represent the best performance, while underlined values denote the second-best performance.}
		
		\renewcommand{\arraystretch}{2}
		\resizebox{2\columnwidth}{!}{
			\begin{tabular}{c|cccc}
				\hline
				Models                & Acuracy & Precision & Recall  & F1-score \\ \hline
				Yadav et al. (VGG16 as backbone) \cite{yadav2019deep} &88.50\% &-&-&-\\
				Ayan et al. (VGG16 as backbone) \cite{ayan2019diagnosis} &87.98\%&82.72\%&85.90\%&84.28\%\\
				Chattopadhyay et al. \cite{chattopadhyay2023exploring}& 81.7\%&-&-&80.6\% \\
				Bhatt et al. (CNN) \cite{bhatt2023convolutional} & 85.58\%&83.33\%&96.15\%&89.29\%\\
				Reshan et al. (MobileNet as backbone) \cite{reshan2023detection}&  90.85\% & 91.41\% &95.28\% & 91.41\%    \\
				Reshan et al. (ResNet152B2 as backbone) \cite{reshan2023detection}&  84.65\% &  82.38\% & 99.21\% & 90.02\%    \\
				Reshan et al. (DenseNet121 as backbone) \cite{reshan2023detection}&  88.90\% &  88.33\% & \textbf{96.87\%} &  92.41\%    \\
				Reshan et al. (Xception as backbone) \cite{reshan2023detection}&  87.59\% & 91.75\% & 90.32\% &  91.03\%    \\
				Reshan et al. (EfficientNet as backbone) \cite{reshan2023detection}&   51.02\% &  86.21\% & 45.85\% &  90.10\%    \\
				Jiang et al. (MP-ViT) \cite{jiang2022multisemantic} & 91.19\% & 91.82\% & 89.36\% & 90.34\% \\
				ViT in \cite{mabrouk2022pneumonia}&  \underline{92.45\%} &  \underline{92.47\%}  & 92.44\%& 92.47\% \\ 
				Proposed method (ResNet-50 as backbone) &91.67\%$\pm$0.6  &92.04\%$\pm$0.5 &94.87\%$\pm$0.5 &\underline{93.43\%}$\pm$0.6 \\
				Proposed method (ResNet-101 as backbone) &\textbf{93.75\%}$\pm$0.5 &\textbf{93.98\%}$\pm$0.5 &\underline{96.16\%}$\pm$0.5 &\textbf{95.05\%}$\pm$0.5 \\ \hline
			\end{tabular}			
		}
		\label{tb:2}
	\end{table*}
	
	\begin{table*}
		\centering
		\caption{Effect of segmentation on "Cohen" dataset}
		
		\renewcommand{\arraystretch}{1.5}
		\resizebox{2\columnwidth}{!}{
			\begin{tabular}{c|c|cccc}
				\hline
				Backbones                  & Results of proposed method & Acuracy & Precision & Recall  & F1-score      \\ \hline
				\multirow{2}{*}{ResNet-50} & on original images         &  91.23\% &  90.94\% & 85.60\% &  88.19\%  \\
				& on predicted masks          &  \textbf{96.04\%}   & \textbf{96.70\%}  & \textbf{94.90\%}    & \textbf{95.77\%}  \\ \hline
				\multirow{2}{*}{ReNet101}  & on original images         & 90.22\% &  88.16\%  & 87.04\% & 87.6\%   \\
				& on predicted masks          &  \textbf{95.19\%}   &  \textbf{95.59\%}  & \textbf{94.19\%}    &   \textbf{94.86\%}    \\ \hline
			\end{tabular}	
		}
		\label{tb:3}
	\end{table*}
	
	\subsubsection{Segmentation}
	
	Our lightweight transformer-enhanced TransUNet, initially designed to improve the accuracy of pneumonia classification by providing precise lung segmentation, also demonstrates superior performance in the segmentation task itself. While the primary goal of our segmentation model was to isolate lung regions for better classification, as shown in Table \ref{tb:seg}, its ability to outperform state-of-the-art segmentation methods highlights the effectiveness of our design choices. By integrating transformer-based attention mechanisms into the U-Net architecture, our model captures both local fine-grained details and global contextual information, which are essential for accurate lung segmentation. To ensure the robustness and reliability of our results, we trained and tested the model three times, with consistent performance across all runs. This dual capability—enhancing both segmentation and classification—sets our approach apart from existing methods, which often focus on one task at the expense of the other.

	\subsubsection{Classification}

	We evaluated the performance of our proposed method on the "Cohen" and "Kermany" datasets, comparing it against several state-of-the-art methods. As shown in Table \ref{tb:1} and Table \ref{tb:2}, our approach outperformed existing methods across all evaluation metrics, including accuracy, precision, recall, and F1-score, using both ResNet-50 and ResNet-101 backbones. On the "Cohen" dataset, our method achieved an accuracy of 96.04\% with ResNet-50 and 95.19\% with ResNet-101, surpassing the best-performing existing method (Zhao et al. \cite{zhao2023dcacorrcapsnet}) by a significant margin. Similarly, on the "Kermany" dataset, our method achieved an accuracy of 93.75\% with ResNet-101 and 91.67\% with ResNet-50, outperforming recent transformer-based models like MP-ViT \cite{jiang2022multisemantic} and ViT \cite{mabrouk2022pneumonia}.
	
	The superior performance of our method can be attributed to several key design choices. First, by leveraging pre-trained ResNet models, we extract robust multi-scale feature representations that capture both low-level and high-level patterns in chest X-rays. Freezing the pre-trained layers allows the model to focus on learning task-specific features in the newly added layers, reducing overfitting and improving generalization. Second, CRAM refine the extracted features through dual attention mechanisms, enhancing discriminative local patterns while suppressing irrelevant information. Our lightweight transformer module then enhances the model's ability to capture global contextual information, which is critical for distinguishing subtle pneumonia-related patterns from complex backgrounds. This integration of local and global feature extraction ensures robust performance across diverse datasets.

	Furthermore, our method demonstrates consistent performance across both datasets, highlighting its generalizability and adaptability to different clinical settings. For instance, on the "Cohen" dataset, our method achieved an F1-score of 95.77\%, significantly higher than the 83.48\% reported by Gadza et al. \cite{gazda2021self}. Similarly, on the "Kermany" dataset, our method achieved an F1-score of 95.05\%, outperforming the 92.41\% achieved by Reshan et al. \cite{reshan2023detection} using DenseNet121. These results underscore the effectiveness of our approach in addressing the challenges of pneumonia classification, such as class imbalance, subtle visual indicators, and domain shifts between datasets.
	
	In summary, our proposed method not only achieves state-of-the-art performance but also demonstrates computational efficiency and generalizability, making it a promising solution for real-world clinical applications. The integration of pre-trained ResNet models with a lightweight transformer module provides a robust framework for accurate and efficient pneumonia detection, addressing the limitations of existing methods.

	\begin{figure*}
		\centering
		\begin{subfigure}[b]{0.75\columnwidth}
			\centering
			\includegraphics[width=0.75\linewidth]{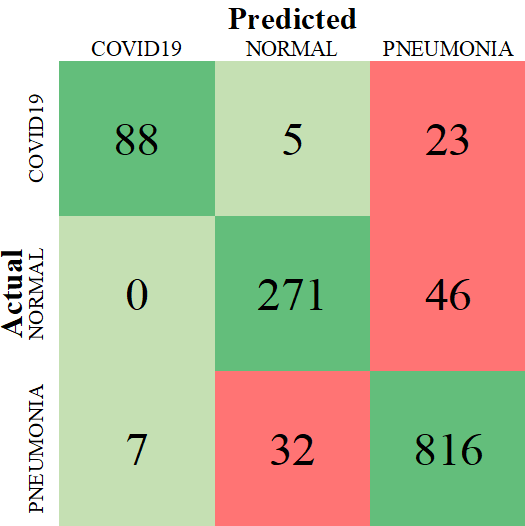}
			\caption{Result of the proposed method on ResNet-50 with original images.}
			\label{res50_org}
		\end{subfigure}
		\begin{subfigure}[b]{0.75\columnwidth}
			\centering
			\includegraphics[width=0.75\linewidth]{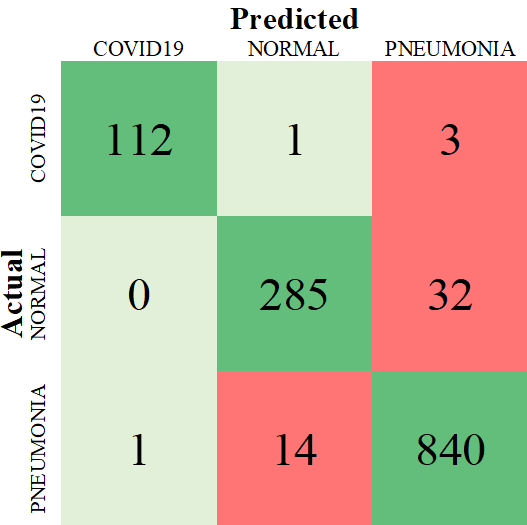}
			\caption{Result of the proposed method on ResNet-50 with predicted masks.}
			\label{res50_mask}
		\end{subfigure}
		\begin{subfigure}[b]{0.75\columnwidth}
			\centering
			\includegraphics[width=0.75\linewidth]{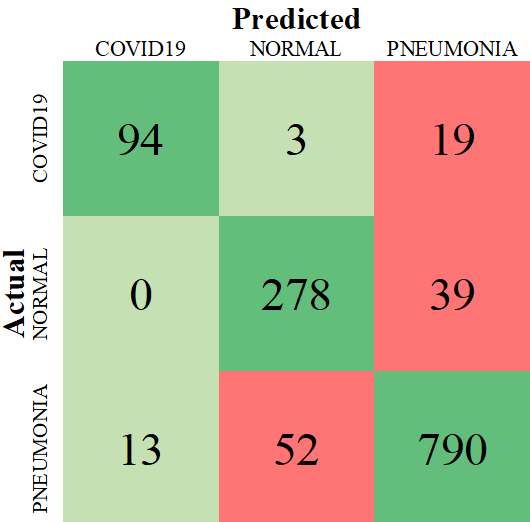}
			\caption{Result of the proposed method on ResNet-101 with original images.}
			\label{res101_org}
		\end{subfigure}
		\begin{subfigure}[b]{0.75\columnwidth}
			\centering
			\includegraphics[width=0.75\linewidth]{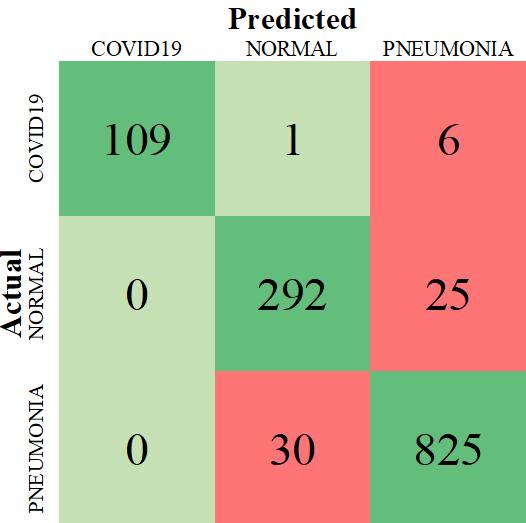}
			\caption{Result of the proposed method on ResNet-101 with predicted masks.}
			\label{res101_mask}
		\end{subfigure}
		
		\caption{Confusion matrix of "Cohen" dataset with different scenarios}
		
		\label{fig:conf}
	\end{figure*}
	\subsection{Ablation study}
	To evaluate the impact of the segmentation component on the performance of our classification model, we conducted an ablation study using the "Cohen" dataset. This study compares the classification results obtained with and without the segmentation step, providing insights into the effectiveness of incorporating lung masks generated by the TransUNet model.
	
	The ablation study involves evaluating the classification performance of our model with two different input scenarios: 1) Original Images: Classification is performed directly on the raw chest X-ray images from the "Cohen" dataset. 2) Predicted Masks: Classification is performed on the chest X-ray images after segmenting the lung regions using the TransUNet model. The images used for classification are limited to the areas highlighted by the predicted lung masks.
	
	The results of the ablation study are summarized in Table \ref{tb:3}. The table displays classification metrics, including accuracy, precision, recall, and F1-score, for both ResNet-50 and ResNet-101 backbones under the two different input scenarios.
	\begin{table*}[h]
		\centering
		\caption{The effect of using each key component on "Cohen" dataset (on original images and without segmentation)}
		
		\renewcommand{\arraystretch}{1.5}
		\resizebox{2\columnwidth}{!}{
			\begin{tabular}{c|ccc|cccc|cc}
				\hline
				Backbones & Baseline & \begin{tabular}[c]{@{}c@{}}Multi-scale\\ feature maps\end{tabular} & \begin{tabular}[c]{@{}c@{}}CRAM +\\ Transformer\end{tabular} & Accuracy & Precision & Recall & F1-score & \begin{tabular}[c]{@{}c@{}}Training time\\(per image)\end{tabular} & \begin{tabular}[c]{@{}c@{}}Learnable\\parameters\end{tabular} \\ \hline
				\multirow{4}{*}{ResNet-50}  & X &   &   & 84.62\% & 75.68\% & 75.10\% & 75.39\% & 0.006s & 0.65 M \\
				& X & X &   & 87.73\% & 80.55\% & 79.62\% & 80.08\% & 0.009s & 0.85 M \\
				& X &   & X & 87.73\% & 80.63\% & 80.47\% & 80.55\% & 0.012s & 1.15 M \\
				& X & X & X & \textbf{91.23\%} & \textbf{90.94\%} & \textbf{85.60\%} & \textbf{88.19\%} & 0.016s & 2.29 M \\ \hline
				\multirow{4}{*}{ResNet-101} & X &   &   & 83.93\% & 74.21\% & 73.92\% & 74.06\% & 0.007s & 0.65 M \\
				& X & X &   & 85.56\% & 78.21\% & 80.4\%  & 79.29\% & 0.011s & 0.85 M \\
				& X &   & X & 85.71\% & 78.59\% & 81.11\% & 79.83\% & 0.016s & 1.15 M \\
				& X & X & X & \textbf{90.22\%} & \textbf{88.16\%} & \textbf{87.04\%} & \textbf{87.6\%}  & 0.027s & 2.29 M \\ \hline
			\end{tabular}
		}
		\label{tb:4}
	\end{table*}

	\begin{table*}[h]
		\centering
		
		\caption{The effect of using different ResNet blocks on "Cohen" dataset}
		
		\renewcommand{\arraystretch}{1.5}
		\resizebox{2\columnwidth}{!}{
			\begin{tabular}{c|ccc|cccc}
				\hline
				Backbones & Block 2 & Block 3 & Block 4 & Accuracy & Precision & Recall & F1-score \\ \hline
				\multirow{5}{*}{ResNet-50} & & & X & 84.62\% & 75.68\% & 75.10\% & 75.39\% \\
				& X & & X & 85.86\% & 77.03\% & 74.37\% & 75.68\% \\
				& & X & X & 86.64\% & 79.36\% & 76.15\% & 77.72\% \\
				& X & X & X & 86.72\% & 79.02\% & 76.48\% & 77.73\% \\
				& \multicolumn{2}{c}{Combined} & X & \textbf{87.73\%} & \textbf{80.55\%} & \textbf{79.62\%} & \textbf{80.08\%}
				\\ \hline
			\end{tabular}
		}
		\label{tb:5}
	\end{table*}

	The results clearly demonstrate the benefit of incorporating segmentation masks in the classification process. For both ResNet-50 and ResNet-101 backbones, the model achieves higher accuracy, precision, recall, and F1-score when trained on images with predicted lung masks compared to the raw images. Specifically, the accuracy improves by almost five percentage points for both ResNet-50 and ResNet-101. Similarly, the precision, recall, and F1-score all show substantial improvements.
	
	These findings underscore the effectiveness of the segmentation component in isolating relevant features within the lung regions, which enhances the classification model's ability to accurately diagnose pneumonia. By focusing on the segmented lung areas, the classification model benefits from reduced noise and more relevant information, leading to better overall performance.
	
	In addition to the classification metrics presented in Table \ref{tb:3}, we further analyze the performance of our model using confusion matrices under different scenarios, as depicted in Figure \ref{fig:conf}. The confusion matrices provide a detailed breakdown of the classification results, showing the true positives, true negatives, false positives, and false negatives for each class. 
	
	For the ResNet-50 backbone, the confusion matrix in Figure \ref{res50_org} shows the results on original images. The model correctly classifies 88 COVID-19 cases, 271 normal cases, and 816 pneumonia cases, with a notable number of misclassifications, particularly in the COVID-19 and pneumonia categories. When using predicted masks, as shown in Figure \ref{res50_mask}, the model's performance improves significantly, correctly classifying 112 COVID-19 cases, 285 normal cases, and 840 pneumonia cases. The number of misclassifications decreases across all categories, highlighting the benefit of segmentation in isolating relevant features.
	
	For the ResNet-101 backbone, the confusion matrix in Figure \ref{res101_org} displays the results on original images, where the model correctly classifies 94 COVID-19 cases, 278 normal cases, and 790 pneumonia cases. However, the misclassifications are more pronounced compared to ResNet-50, particularly in the pneumonia category. With the predicted masks, as shown in Figure \ref{res101_mask}, the performance improves, with correct classifications of 109 COVID-19 cases, 292 normal cases, and 825 pneumonia cases. This reduction in misclassifications further supports the effectiveness of incorporating segmentation masks.
	
	These visual representations in the confusion matrices clearly demonstrate the improvement in classification performance when using the predicted masks. The consistent reduction in false positives and false negatives across both backbones underscores the robustness of the segmentation approach. By focusing on the lung regions and eliminating irrelevant background information, the segmentation component enhances the model's ability to accurately diagnose pneumonia, resulting in better overall performance.

	Furthermore, we investigate the contributions of key components within our model on the "Cohen" dataset, specifically focusing on the impact of multi-scale feature maps and the transformer module. The results are presented in Table \ref{tb:4}, highlighting the performance metrics, including accuracy, precision, recall, and F1-score, across different configurations of the ResNet-50 and ResNet-101 backbones. Each row of the table demonstrates how the incorporation of each component affects the model's performance, providing a comprehensive view of their individual and combined effects.

	The findings reveal a significant enhancement in model performance when both multi-scale feature maps and the transformer are employed alongside the baseline configuration. For instance, with the ResNet-50 backbone, the accuracy improves from 84.62\% (baseline) to 91.23\% when all components are utilized. Similarly, the ResNet-101 backbone exhibits a notable increase in accuracy from 83.93\% to 90.22\%. These results underscore the effectiveness of our proposed innovations, illustrating that the integration of multi-scale feature maps and transformer elements not only enhances overall accuracy but also boosts precision, recall, and F1-score, which are crucial for the reliability of classification tasks. This highlights the importance of these key components in achieving improved performance in deep learning models for image analysis. Additionally, the inclusion of training time (per image) and learnable parameters underscores the computational efficiency and scalability of our approach, ensuring its suitability for resource-constrained environments.

	\begin{figure}[h]
		\centering
		\includegraphics[width=0.95\linewidth]{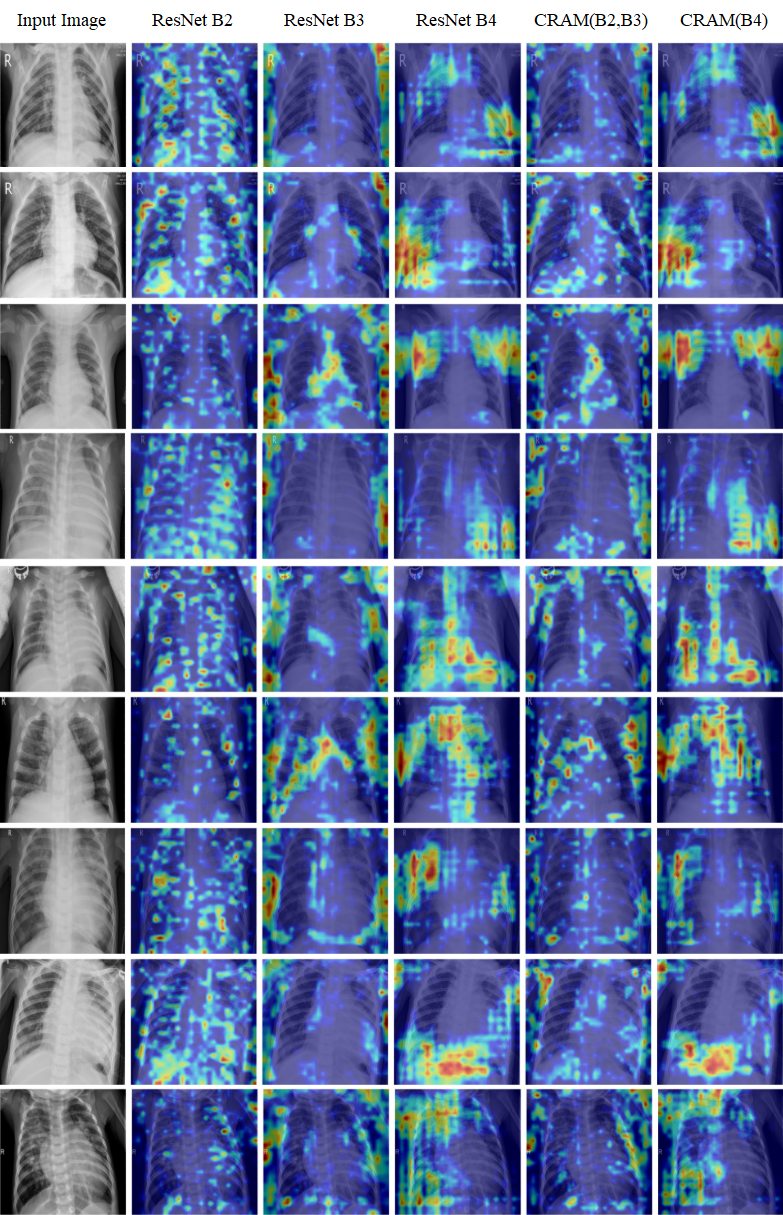}
		
		\caption{Grad-CAM visualizations on Kermany dataset samples}

		\label{fig:gradcam}
	\end{figure}

	Additionally, we conducted an ablation study to evaluate the contribution of each ResNet block (Blocks 2, 3, and 4) to the classification performance. The results, as shown in Table \ref{tb:5}, demonstrate that intermediate layers (Blocks 2 and 3) play a significant role in improving performance, but their contribution is not as critical as that of Block 4. For instance, using only Block 4 achieves an accuracy of 84.62\%, while adding Block 3 (without Block 2) improves the accuracy to 86.64\%. This indicates that mid-level features from Block 3 provide additional discriminative information, enhancing the model's ability to detect subtle patterns in chest X-rays. However, the inclusion of Block 2, which captures low-level features, results in only a marginal improvement in accuracy (86.72\%), suggesting that while low-level features are beneficial, they are less impactful compared to mid- and high-level features.
	
	Interestingly, the best performance is achieved when using two feature maps: Block 4 ($B^4$) and the merged feature map ($B^{merged}$), which combines Blocks 2 and 3. This configuration achieves an accuracy of 87.73\%, a precision of 80.55\%, and an F1-score of 80.08\%, outperforming the scenario where all three blocks are used independently. This result highlights the importance of separating high-level features (Block 4) from intermediate features (Blocks 2 and 3). While intermediate layers provide valuable contextual information, they are not as discriminative as the high-level features from Block 4. By merging Blocks 2 and 3 into a single feature map ($B^{merged}$), we reduce redundancy and computational complexity while preserving the benefits of multi-scale feature extraction. This design choice ensures that the model focuses on the most relevant features for pneumonia detection, leading to improved performance and efficiency.

	\subsection{Explainable AI through Gradient-Weighted Class Activation Mapping}

	To enhance the interpretability of our model and provide clinical insights into its decision-making process, we employed Gradient-Weighted Class Activation Mapping (Grad-CAM) visualizations. Explainable AI is particularly crucial in medical diagnostics, where understanding the rationale behind a model's predictions is essential for clinical adoption and trust. Grad-CAM generates heatmaps that highlight the regions of the input image that most significantly influenced the model's classification decision, effectively creating a visual explanation for the prediction.

	As shown in Figure \ref{fig:gradcam}, our Grad-CAM analysis reveals the hierarchical feature learning process across different network depths. The first column displays the original input chest X-ray image, providing the baseline for comparison. The subsequent three columns visualize the activation maps from the feature maps extracted at different ResNet blocks: Block 2 (capturing low-level features and basic patterns), Block 3 (capturing mid-level features and structural information), and Block 4 (capturing high-level semantic features and complex patterns). 	The final two columns present the activation maps from our CRAM: the concatenated feature map (merging Blocks 2 and 3) and the enhanced Block 4 feature map.
	
	
	

	\section{Conclusion}
	\label{conclu}
	This paper presents an innovative and efficient method for pneumonia detection utilizing a novel multi-scale transformer approach. By integrating lung segmentation using the TransUNet model with a specialized transformer module, our approach effectively isolates lung regions, thereby enhancing the performance of subsequent classification tasks. The proposed method demonstrates significant improvements in classification metrics, as evidenced by the ablation study on the "Cohen" dataset. Both ResNet-50 and ResNet-101 backbones benefited from the segmentation masks, showing increased accuracy, precision, recall, and F1-score. These improvements underscore the effectiveness of our approach in focusing on relevant lung features while reducing noise from irrelevant regions.
	
	The high accuracy rates of 93.75\% on the "Kermany" dataset and 96.04\% on the "Cohen" dataset confirm the robustness and reliability of our model. The reduction in the number of parameters compared to other state-of-the-art transformer models highlights our contribution to creating a more efficient yet powerful diagnostic tool suitable for deployment in resource-constrained environments. Our work paves the way for future research in several areas. Future work could explore further optimization of the transformer module to enhance performance and reduce computational complexity. Additionally, expanding the dataset to include a broader variety of pneumonia cases and other respiratory diseases could improve the model's generalization. 
	\bibliographystyle{IEEEtran}
	\bibliography{ref1.bib}

\begin{thebibliography}{10}
\providecommand{\url}[1]{#1}
\csname url@samestyle\endcsname
\providecommand{\newblock}{\relax}
\providecommand{\bibinfo}[2]{#2}
\providecommand{\BIBentrySTDinterwordspacing}{\spaceskip=0pt\relax}
\providecommand{\BIBentryALTinterwordstretchfactor}{4}
\providecommand{\BIBentryALTinterwordspacing}{\spaceskip=\fontdimen2\font plus
\BIBentryALTinterwordstretchfactor\fontdimen3\font minus
  \fontdimen4\font\relax}
\providecommand{\BIBforeignlanguage}[2]{{%
\expandafter\ifx\csname l@#1\endcsname\relax
\typeout{** WARNING: IEEEtran.bst: No hyphenation pattern has been}%
\typeout{** loaded for the language `#1'. Using the pattern for}%
\typeout{** the default language instead.}%
\else
\language=\csname l@#1\endcsname
\fi
#2}}
\providecommand{\BIBdecl}{\relax}
\BIBdecl

\bibitem{unicef_pneumonia}
UNICEF, ``Pneumonia,''
  \url{https://data.unicef.org/topic/child-health/pneumonia/}, last update:
  2023-11.

\bibitem{who_pneumonia_in_children}
W.~H. Organization, ``Pneumonia in children,''
  \url{https://www.who.int/news-room/fact-sheets/detail/pneumonia}, last
  update: 2022-11.

\bibitem{health_poverty_action_poverty_and_health}
H.~P. Action, ``Key facts: Poverty and poor health,''
  \url{https://www.healthpovertyaction.org/news-events/key-facts-poverty-and-poor-health/},
  last update: 2018-01.

\bibitem{radiologyinfo_chest_xray}
RadiologyInfo, ``Chest x-ray,''
  \url{https://www.radiologyinfo.org/en/info/chestrad}, last update: 2022-11.

\bibitem{lin2023enhancing}
M.~Lin, B.~Hou, S.~Mishra, T.~Yao, Y.~Huo, Q.~Yang, F.~Wang, G.~Shih, and
  Y.~Peng, ``Enhancing thoracic disease detection using chest x-rays from
  pubmed central open access,'' \emph{Computers in biology and medicine}, vol.
  159, p. 106962, 2023.

\bibitem{askari2025enhancing}
F.~Askari, A.~Fateh, and M.~R. Mohammadi, ``Enhancing few-shot image
  classification through learnable multi-scale embedding and attention
  mechanisms,'' \emph{Neural Networks}, p. 107339, 2025.

\bibitem{fateh2024msdnet}
A.~Fateh, M.~R. Mohammadi, and M.~R.~J. Motlagh, ``Msdnet: Multi-scale decoder
  for few-shot semantic segmentation via transformer-guided prototyping,''
  \emph{arXiv preprint arXiv:2409.11316}, 2024.

\bibitem{song2025multi}
Z.~Song, W.~Wu, and S.~Wu, ``Multi-scale convolutional attention and structural
  re-parameterized residual-based 3d u-net for liver and liver tumor
  segmentation from ct,'' \emph{Sensors (Basel, Switzerland)}, vol.~25, no.~6,
  p. 1814, 2025.

\bibitem{junia2024deep}
R.~C. Junia and K.~Selvan, ``Deep learning-based automatic segmentation of
  covid-19 in chest x-ray images using ensemble neural net sentinel
  algorithm,'' \emph{Measurement: Sensors}, vol.~33, p. 101117, 2024.

\bibitem{shavkatovich2025binary}
A.~Shavkatovich~Buriboev, A.~Abduvaitov, and H.~S. Jeon, ``Binary
  classification of pneumonia in chest x-ray images using modified
  contrast-limited adaptive histogram equalization algorithm,'' \emph{Sensors},
  vol.~25, no.~13, p. 3976, 2025.

\bibitem{buriboev2024concatenated}
A.~S. Buriboev, D.~Muhamediyeva, H.~Primova, D.~Sultanov, K.~Tashev, and H.~S.
  Jeon, ``Concatenated cnn-based pneumonia detection using a fuzzy-enhanced
  dataset,'' \emph{Sensors}, vol.~24, no.~20, p. 6750, 2024.

\bibitem{kanwal2024current}
K.~Kanwal, M.~Asif, S.~G. Khalid, H.~Liu, A.~G. Qurashi, and S.~Abdullah,
  ``Current diagnostic techniques for pneumonia: A scoping review,''
  \emph{Sensors}, vol.~24, no.~13, p. 4291, 2024.

\bibitem{candemir2013lung}
S.~Candemir, S.~Jaeger, K.~Palaniappan, J.~P. Musco, R.~K. Singh, Z.~Xue,
  A.~Karargyris, S.~Antani, G.~Thoma, and C.~J. McDonald, ``Lung segmentation
  in chest radiographs using anatomical atlases with nonrigid registration,''
  \emph{IEEE transactions on medical imaging}, vol.~33, no.~2, pp. 577--590,
  2013.

\bibitem{jaeger2013automatic}
S.~Jaeger, A.~Karargyris, S.~Candemir, L.~Folio, J.~Siegelman, F.~Callaghan,
  Z.~Xue, K.~Palaniappan, R.~K. Singh, S.~Antani \emph{et~al.}, ``Automatic
  tuberculosis screening using chest radiographs,'' \emph{IEEE transactions on
  medical imaging}, vol.~33, no.~2, pp. 233--245, 2013.

\bibitem{kermany2018labeled}
D.~Kermany, K.~Zhang, M.~Goldbaum \emph{et~al.}, ``Labeled optical coherence
  tomography (oct) and chest x-ray images for classification,'' \emph{Mendeley
  data}, vol.~2, no.~2, p. 651, 2018.

\bibitem{cohen2020covid}
\BIBentryALTinterwordspacing
J.~P. Cohen, P.~Morrison, and L.~Dao, ``Covid-19 image data collection,''
  \emph{arXiv 2003.11597}, 2020. [Online]. Available:
  \url{https://github.com/ieee8023/covid-chestxray-dataset}
\BIBentrySTDinterwordspacing

\bibitem{jennifer2023neutrosophic}
J.~S. Jennifer and T.~S. Sharmila, ``A neutrosophic set approach on chest
  x-rays for automatic lung infection detection,'' \emph{Information Technology
  and Control}, vol.~52, no.~1, pp. 37--52, 2023.

\bibitem{fateh2023persian}
A.~Fateh, M.~Rezvani, A.~Tajary, and M.~Fateh, ``Persian printed text line
  detection based on font size,'' \emph{Multimedia Tools and Applications},
  vol.~82, no.~2, pp. 2393--2418, 2023.

\bibitem{fateh2022providing}
------, ``Providing a voting-based method for combining deep neural network
  outputs to layout analysis of printed documents,'' \emph{Journal of Machine
  Vision and Image Processing}, vol.~9, no.~1, pp. 47--64, 2022.

\bibitem{ronneberger2015u}
O.~Ronneberger, P.~Fischer, and T.~Brox, ``U-net: Convolutional networks for
  biomedical image segmentation,'' in \emph{Medical image computing and
  computer-assisted intervention--MICCAI 2015: 18th international conference,
  Munich, Germany, October 5-9, 2015, proceedings, part III 18}.\hskip 1em plus
  0.5em minus 0.4em\relax Springer, 2015, pp. 234--241.

\bibitem{liu2022automatic}
W.~Liu, J.~Luo, Y.~Yang, W.~Wang, J.~Deng, and L.~Yu, ``Automatic lung
  segmentation in chest x-ray images using improved u-net,'' \emph{Scientific
  Reports}, vol.~12, no.~1, p. 8649, 2022.

\bibitem{huang2020unet}
H.~Huang, L.~Lin, R.~Tong, H.~Hu, Q.~Zhang, Y.~Iwamoto, X.~Han, Y.-W. Chen, and
  J.~Wu, ``Unet 3+: A full-scale connected unet for medical image
  segmentation,'' in \emph{ICASSP 2020-2020 IEEE international conference on
  acoustics, speech and signal processing (ICASSP)}.\hskip 1em plus 0.5em minus
  0.4em\relax IEEE, 2020, pp. 1055--1059.

\bibitem{jha2020doubleu}
D.~Jha, M.~A. Riegler, D.~Johansen, P.~Halvorsen, and H.~D. Johansen,
  ``Doubleu-net: A deep convolutional neural network for medical image
  segmentation,'' in \emph{2020 IEEE 33rd International symposium on
  computer-based medical systems (CBMS)}.\hskip 1em plus 0.5em minus
  0.4em\relax IEEE, 2020, pp. 558--564.

\bibitem{oktay2018attention}
O.~Oktay, J.~Schlemper, L.~L. Folgoc, M.~Lee, M.~Heinrich, K.~Misawa, K.~Mori,
  S.~McDonagh, N.~Y. Hammerla, B.~Kainz \emph{et~al.}, ``Attention u-net:
  Learning where to look for the pancreas,'' \emph{arXiv preprint
  arXiv:1804.03999}, 2018.

\bibitem{chen2021transunet}
J.~Chen, Y.~Lu, Q.~Yu, X.~Luo, E.~Adeli, Y.~Wang, L.~Lu, A.~L. Yuille, and
  Y.~Zhou, ``Transunet: Transformers make strong encoders for medical image
  segmentation,'' \emph{arXiv preprint arXiv:2102.04306}, 2021.

\bibitem{chen2023cotrfuse}
Y.~Chen, T.~Wang, H.~Tang, L.~Zhao, X.~Zhang, T.~Tan, Q.~Gao, M.~Du, and
  T.~Tong, ``Cotrfuse: a novel framework by fusing cnn and transformer for
  medical image segmentation,'' \emph{Physics in Medicine \& Biology}, vol.~68,
  no.~17, p. 175027, 2023.

\bibitem{stokes2021machine}
K.~Stokes, R.~Castaldo, M.~Franzese, M.~Salvatore, G.~Fico, L.~G. Pokvic,
  A.~Badnjevic, and L.~Pecchia, ``A machine learning model for supporting
  symptom-based referral and diagnosis of bronchitis and pneumonia in limited
  resource settings,'' \emph{Biocybernetics and biomedical engineering},
  vol.~41, no.~4, pp. 1288--1302, 2021.

\bibitem{chandra2020pneumonia}
T.~B. Chandra and K.~Verma, ``Pneumonia detection on chest x-ray using machine
  learning paradigm,'' in \emph{Proceedings of 3rd International Conference on
  Computer Vision and Image Processing: CVIP 2018, Volume 1}.\hskip 1em plus
  0.5em minus 0.4em\relax Springer, 2020, pp. 21--33.

\bibitem{wang2024prediction}
Y.~Wang, Z.-L. Liu, H.~Yang, R.~Li, S.-J. Liao, Y.~Huang, M.-H. Peng, X.~Liu,
  G.-Y. Si, Q.-Z. He \emph{et~al.}, ``Prediction of viral pneumonia based on
  machine learning models analyzing pulmonary inflammation index scores,''
  \emph{Computers in Biology and Medicine}, vol. 169, p. 107905, 2024.

\bibitem{fateh2021multilingual}
A.~Fateh, M.~Fateh, and V.~Abolghasemi, ``Multilingual handwritten numeral
  recognition using a robust deep network joint with transfer learning,''
  \emph{Information Sciences}, vol. 581, pp. 479--494, 2021.

\bibitem{allioui2022multi}
H.~Allioui, M.~A. Mohammed, N.~Benameur, B.~Al-Khateeb, K.~H. Abdulkareem,
  B.~Garcia-Zapirain, R.~Dama{\v{s}}evi{\v{c}}ius, and R.~Maskeli{\=u}nas, ``A
  multi-agent deep reinforcement learning approach for enhancement of covid-19
  ct image segmentation,'' \emph{Journal of personalized medicine}, vol.~12,
  no.~2, p. 309, 2022.

\bibitem{stephen2019efficient}
O.~Stephen, M.~Sain, U.~J. Maduh, and D.-U. Jeong, ``An efficient deep learning
  approach to pneumonia classification in healthcare,'' \emph{Journal of
  healthcare engineering}, vol. 2019, no.~1, p. 4180949, 2019.

\bibitem{rajpurkar2017chexnet}
P.~Rajpurkar, J.~Irvin, K.~Zhu, B.~Yang, H.~Mehta, T.~Duan, D.~Ding, A.~Bagul,
  C.~Langlotz, K.~Shpanskaya \emph{et~al.}, ``Chexnet: Radiologist-level
  pneumonia detection on chest x-rays with deep learning,'' \emph{arXiv
  preprint arXiv:1711.05225}, 2017.

\bibitem{ukwuoma2023hybrid}
C.~C. Ukwuoma, Z.~Qin, M.~B.~B. Heyat, F.~Akhtar, O.~Bamisile, A.~Y. Muaad,
  D.~Addo, and M.~A. Al-Antari, ``A hybrid explainable ensemble transformer
  encoder for pneumonia identification from chest x-ray images,'' \emph{Journal
  of Advanced Research}, vol.~48, pp. 191--211, 2023.

\bibitem{jaiswal2019identifying}
A.~K. Jaiswal, P.~Tiwari, S.~Kumar, D.~Gupta, A.~Khanna, and J.~J. Rodrigues,
  ``Identifying pneumonia in chest x-rays: A deep learning approach,''
  \emph{Measurement}, vol. 145, pp. 511--518, 2019.

\bibitem{gabruseva2020deep}
T.~Gabruseva, D.~Poplavskiy, and A.~Kalinin, ``Deep learning for automatic
  pneumonia detection,'' in \emph{Proceedings of the IEEE/CVF conference on
  computer vision and pattern recognition workshops}, 2020, pp. 350--351.

\bibitem{Gholamiije25}
M.~Gholami, M.~Fateh, and A.~Fateh, ``Text-enhanced semantic segmentation via
  contrastive language-image pretraining guided multi-modal feature fusion with
  feature refinement approach,'' \emph{International Journal of Engineering},
  vol.~39, no.~6, pp. 1422--1437, 2026.

\bibitem{wang2021transpath}
X.~Wang, S.~Yang, J.~Zhang, M.~Wang, J.~Zhang, J.~Huang, W.~Yang, and X.~Han,
  ``Transpath: Transformer-based self-supervised learning for histopathological
  image classification,'' in \emph{Medical Image Computing and Computer
  Assisted Intervention--MICCAI 2021: 24th International Conference,
  Strasbourg, France, September 27--October 1, 2021, Proceedings, Part VIII
  24}.\hskip 1em plus 0.5em minus 0.4em\relax Springer, 2021, pp. 186--195.

\bibitem{wu2022swin}
P.~Wu, J.~Chen, and Y.~Wu, ``Swin transformer based benign and malignant
  pulmonary nodule classification,'' in \emph{5th International Conference on
  Computer Information Science and Application Technology (CISAT 2022)}, vol.
  12451.\hskip 1em plus 0.5em minus 0.4em\relax SPIE, 2022, pp. 552--558.

\bibitem{liu2022swin}
Z.~Liu, H.~Hu, Y.~Lin, Z.~Yao, Z.~Xie, Y.~Wei, J.~Ning, Y.~Cao, Z.~Zhang,
  L.~Dong \emph{et~al.}, ``Swin transformer v2: Scaling up capacity and
  resolution,'' in \emph{Proceedings of the IEEE/CVF conference on computer
  vision and pattern recognition}, 2022, pp. 12\,009--12\,019.

\bibitem{mishra2024empowering}
V.~R. Mishra and R.~Malhotra, ``Empowering healthcare with swin transformer v2:
  Advancing pneumonia diagnosis through deep learning,'' in \emph{Computational
  Methods in Science and Technology}.\hskip 1em plus 0.5em minus 0.4em\relax
  CRC Press, 2024, pp. 99--105.

\bibitem{angara2024novel}
S.~Angara, N.~R. Mannuru, A.~Mannuru, and S.~Thirunagaru, ``A novel method to
  enhance pneumonia detection via a model-level ensembling of cnn and vision
  transformer,'' \emph{arXiv preprint arXiv:2401.02358}, 2024.

\bibitem{mustapha2025enhanced}
B.~Mustapha, Y.~Zhou, C.~Shan, and Z.~Xiao, ``Enhanced pneumonia detection in
  chest x-rays using hybrid convolutional and vision transformer networks,''
  \emph{Current Medical Imaging}, p. e15734056326685, 2025.

\bibitem{anbalagan2023analysis}
T.~Anbalagan, M.~K. Nath, D.~Vijayalakshmi, and A.~Anbalagan, ``Analysis of
  various techniques for ecg signal in healthcare, past, present, and future,''
  \emph{Biomedical Engineering Advances}, vol.~6, p. 100089, 2023.

\bibitem{alom2018nuclei}
M.~Z. Alom, C.~Yakopcic, T.~M. Taha, and V.~K. Asari, ``Nuclei segmentation
  with recurrent residual convolutional neural networks based u-net
  (r2u-net),'' in \emph{NAECON 2018-IEEE National Aerospace and Electronics
  Conference}.\hskip 1em plus 0.5em minus 0.4em\relax IEEE, 2018, pp. 228--233.

\bibitem{lau2020automated}
S.~L. Lau, E.~K. Chong, X.~Yang, and X.~Wang, ``Automated pavement crack
  segmentation using u-net-based convolutional neural network,'' \emph{Ieee
  Access}, vol.~8, pp. 114\,892--114\,899, 2020.

\bibitem{azad2019bi}
R.~Azad, M.~Asadi-Aghbolaghi, M.~Fathy, and S.~Escalera, ``Bi-directional
  convlstm u-net with densley connected convolutions,'' in \emph{Proceedings of
  the IEEE/CVF international conference on computer vision workshops}, 2019,
  pp. 0--0.

\bibitem{jalali2021resbcdu}
Y.~Jalali, M.~Fateh, M.~Rezvani, V.~Abolghasemi, and M.~H. Anisi,
  ``Resbcdu-net: a deep learning framework for lung ct image segmentation,''
  \emph{Sensors}, vol.~21, no.~1, p. 268, 2021.

\bibitem{zhang2022automatic}
Y.~Zhang, B.~D. Davison, V.~W. Talghader, Z.~Chen, Z.~Xiao, and G.~J. Kunkel,
  ``Automatic head overcoat thickness measure with nasnet-large-decoder net,''
  in \emph{Proceedings of the Future Technologies Conference (FTC) 2021, Volume
  2}.\hskip 1em plus 0.5em minus 0.4em\relax Springer, 2022, pp. 159--176.

\bibitem{jalali2024dabt}
Y.~Jalali, M.~Fateh, and M.~Rezvani, ``Dabt-u-net: Dual attentive bconvlstm
  u-net with transformers and collaborative patch-based approach for accurate
  retinal vessel segmentation,'' \emph{International Journal of Engineering},
  vol.~37, no.~10, pp. 2051--2065, 2024.

\bibitem{rezvani2024abanet}
S.~Rezvani, M.~Fateh, and H.~Khosravi, ``Abanet: Attention boundary-aware
  network for image segmentation,'' \emph{Expert Systems}, vol.~41, no.~9, p.
  e13625, 2024.

\bibitem{rezvani2025fusionlungnet}
S.~Rezvani, M.~Fateh, Y.~Jalali, and A.~Fateh, ``Fusionlungnet: Multi-scale
  fusion convolution with refinement network for lung ct image segmentation,''
  \emph{Biomedical Signal Processing and Control}, vol. 107, p. 107858, 2025.

\bibitem{goodwin2020intra}
B.~D. Goodwin, C.~Jaskolski, C.~Zhong, and H.~Asmani, ``Intra-model variability
  in covid-19 classification using chest x-ray images,'' \emph{arXiv preprint
  arXiv:2005.02167}, 2020.

\bibitem{gazda2021self}
M.~Gazda, J.~Plavka, J.~Gazda, and P.~Drotar, ``Self-supervised deep
  convolutional neural network for chest x-ray classification,'' \emph{IEEE
  Access}, vol.~9, pp. 151\,972--151\,982, 2021.

\bibitem{zhao2023dcacorrcapsnet}
A.~Zhao, H.~Wu, M.~Chen, and N.~Wang, ``Dcacorrcapsnet: A deep
  channel-attention correlative capsule network for covid-19 detection based on
  multi-source medical images,'' \emph{IET Image Processing}, vol.~17, no.~4,
  pp. 988--1000, 2023.

\bibitem{van2024large}
M.-H. Van, P.~Verma, and X.~Wu, ``On large visual language models for medical
  imaging analysis: An empirical study,'' in \emph{2024 IEEE/ACM Conference on
  Connected Health: Applications, Systems and Engineering Technologies
  (CHASE)}.\hskip 1em plus 0.5em minus 0.4em\relax IEEE, 2024, pp. 172--176.

\bibitem{yadav2019deep}
S.~S. Yadav and S.~M. Jadhav, ``Deep convolutional neural network based medical
  image classification for disease diagnosis,'' \emph{Journal of Big data},
  vol.~6, no.~1, pp. 1--18, 2019.

\bibitem{ayan2019diagnosis}
E.~Ayan and H.~M. {\"U}nver, ``Diagnosis of pneumonia from chest x-ray images
  using deep learning,'' in \emph{2019 Scientific meeting on
  electrical-electronics \& biomedical engineering and computer science
  (EBBT)}.\hskip 1em plus 0.5em minus 0.4em\relax Ieee, 2019, pp. 1--5.

\bibitem{chattopadhyay2023exploring}
S.~Chattopadhyay, S.~Ganguly, S.~Chaudhury, S.~Nag, and S.~Chattopadhyay,
  ``Exploring self-supervised representation learning for low-resource medical
  image analysis,'' in \emph{2023 IEEE International Conference on Image
  Processing (ICIP)}.\hskip 1em plus 0.5em minus 0.4em\relax IEEE, 2023, pp.
  1440--1444.

\bibitem{bhatt2023convolutional}
H.~Bhatt and M.~Shah, ``A convolutional neural network ensemble model for
  pneumonia detection using chest x-ray images,'' \emph{Healthcare Analytics},
  vol.~3, p. 100176, 2023.

\bibitem{reshan2023detection}
M.~S.~A. Reshan, K.~S. Gill, V.~Anand, S.~Gupta, H.~Alshahrani, A.~Sulaiman,
  and A.~Shaikh, ``Detection of pneumonia from chest x-ray images utilizing
  mobilenet model,'' in \emph{Healthcare}, vol.~11, no.~11.\hskip 1em plus
  0.5em minus 0.4em\relax MDPI, 2023, p. 1561.

\bibitem{jiang2022multisemantic}
Z.~Jiang and L.~Chen, ``Multisemantic level patch merger vision transformer for
  diagnosis of pneumonia,'' \emph{Computational and Mathematical Methods in
  Medicine}, vol. 2022, no.~1, p. 7852958, 2022.

\bibitem{mabrouk2022pneumonia}
A.~Mabrouk, R.~P. Diaz~Redondo, A.~Dahou, M.~Abd~Elaziz, and M.~Kayed,
  ``Pneumonia detection on chest x-ray images using ensemble of deep
  convolutional neural networks,'' \emph{Applied Sciences}, vol.~12, no.~13, p.
  6448, 2022.

\end{thebibliography}
	
\end{document}